\documentclass[12pt,modern]{aastex62} 
\usepackage{verbatim}
\usepackage{natbib,aas_macros}
\usepackage{graphicx}
\usepackage{xspace}
\usepackage{xcolor}  
\usepackage{rotating} 

\newcommand{\ho}{$H_{0}$\xspace}                  
   
\newcommand{\hst}{\emph{HST}\xspace}              
    
\newcommand{\hstacs}{\emph{HST/ACS}\xspace}

\newcommand{\sne}{SNe~Ia\xspace}

\newcommand{\ic}{IC\,1613\xspace}

\shorttitle{TRGB Distance Scale}
\shortauthors{Freedman, Madore   et al.}

\begin{document}


\title{\bf Calibration of the Tip of the Red Giant Branch (TRGB) }


\author{\bf Wendy L. Freedman}
\affil{Dept. of Astronomy \& Astrophysics, Univ. Chicago, 5640 S. Ellis Ave., Chicago, IL, ~~60637} 
\email{wfreedman@uchicago.edu}
\author{\bf Barry F. Madore} 
\affil{The Observatories, Carnegie
Institution for Science, 813 Santa Barbara St., Pasadena, CA ~~91101}
\affil{Dept. of Astronomy \& Astrophysics, Univ. Chicago, 5640 S. Ellis Ave., Chicago, IL, ~~60637}
\author{\bf Taylor Hoyt}
\affil{Dept. of Astronomy \& Astrophysics, Univ. Chicago, 5640 S. Ellis Ave., Chicago, IL, ~~60637} 
\author{\bf In Sung Jang}
\affil{Leibniz-Institut f\"{u}r Astrophysik Potsdam (AIP), An der Sternwarte 16, 14482 Potsdam, Germany}
\author{\bf Rachael Beaton}
\affil{Dept. of Astrophysical Sciences, Princeton University, 4 Ivy Lane, Princeton, NJ ~~08544}
\author{\bf Myung Gyoon Lee}
\affil{Department of Physics \& Astronomy, Seoul National University, Gwanak-gu, Seoul 151-742, Korea}
\author{\bf Andrew Monson}
\affil{Dept. of Astronomy \& Astrophysics, Penn State, 525 Davey Lab, University Park, PA 16802}
\author{\bf Jill Neeley}
\affil{Dept. of Physics, Florida Atlantic University, 777 Glades Road, Boca Raton, FL ~~33431}
\author{\bf Jeffrey Rich}
\affil{The Observatories, Carnegie
Institution for Science,
813 Santa Barbara St., Pasadena, CA ~~91101}


\begin{abstract} 

The Tip of the Red Giant (TRGB) method provides one of the most accurate and precise  means of measuring the distances to nearby galaxies. Here we present a  $VIJHK$ absolute calibration of the TRGB based on observations of TRGB stars in the Large Magellanic Cloud (LMC),grounded on detached eclipsing binaries (DEBs). This paper presents a more detailed description of the method first presented in Freedman et al. (2019) for measuring corrections for the total line-of-sight extinction and reddening to the LMC. In this method, we use a differential comparison of the red giant population in the LMC, first with red giants in the Local Group galaxy, IC~1613, and then with those in the Small Magellanic Cloud. As a consistency check, we derive an independent calibration of the TRGB sequence using the SMC alone, invoking its geometric distance also calibrated by DEBs. An additional consistency check comes from near-infrared observations of Galactic globular clusters covering a wide range of metallicities.  In all cases we find excellent agreement in the zero-point calibration. We then examine the recent claims by Yuan et al. (2019), demonstrating that, in the case of the SMC, they corrected for extinction alone while neglecting the essential correction for reddening as well. In the case of IC~1613, we show that their analysis contains an incorrect treatment of (over-correction for) metallicity.   Using our revised (and direct) measurement of the LMC TRGB extinction, we find a value of H$_0$ = 69.6 $\pm$ 0.8 ($\pm$1.1\% stat) $\pm$ 1.7 (±2.4\% sys) km s$^{-1}$ Mpc$^{-1}$. 

\end{abstract}


\section{Introduction}

The Tip of the Red Giant Branch (TRGB) sequence marks the well-understood, abrupt evolutionary transition of low-mass RGB stars onto the  lower-luminosity horizontal branch. These RGB stars are powered by a  hydrogen-burning shell surrounding an isothermal helium core that is supported by electron degeneracy. Their transition away from the TRGB is initiated by a brief helium flash, a period of thermal run-away whereupon the degeneracy of the core is lifted, and the triple-alpha (helium-burning) process can proceed at a lower luminosity in a gas-pressure-supported core. At this transition, the tip stars rapidly fade in luminosity, settling onto the horizontal branch, burning helium in their core and hydrogen in a surrounding shell. The basic underlying physical explanation for a universal upper limit to the luminosity of an RGB star is theoretically well understood (e.g., Salaris et al. 1997, Bildsten et al. 2012;  Serenelli et al. 2017); and empirically, this distinctive feature in the observed RGB luminosity function is well-known to be an excellent means of measuring the distances in the halos of nearby, resolved galaxies: in the $I$ band the TRGB is a `standard candle' (e.g., Lee, Freedman \& Madore 1993; Rizzi et al. 2007; Serenelli et al. 2017). 

Over the past 30 years the TRGB sequence has been  measured and successfully used by many independent groups to determine high-precision distances to large numbers of nearby galaxies (e.g.,  Sakai et al. 2004; Makarov et al. 2006; Rizzi et al. 2007; Dalcanton et al. 2009, 2012; Jang \& Lee 2017a,b; McQuinn et al, 2019; Freedman et al, 2019 and references therein). The unique imaging capabilities of the {\it Hubble Space Telescope} (\hst), combined with the ease with which the $I$-band tip in particular (see below) can be measured and used to determine distances, have made this method readily accessible and widely applied by the general astronomical community. Indeed, a recent search of the literature (NED August 2019) shows that there are currently some 900 applications of the TRGB method published for over 300 individual galaxies. A program aimed at optimally determining TRGB distances simply requires targeted observations of the halos of any nearby galaxy, made to sufficient depth to resolve and measure the brightest  old Population~II red giant stars, which are known to inhabit all stellar halos. Although the method itself is not new, it is only recently that the TRGB method has been applied directly to the determination of H$_o$ (Karachentseva et al. 2003, Mould et al. 2008, 2009; Jang \& Lee 2017; Freedman et al. 2019).

One of the distinct advantages of the TRGB method is the fact that the TRGB stars populate the halos of the host galaxy (in contrast to the Cepheids, for example, which are found only in the higher-surface-brightness disk regions). Additionally, the halos of undisturbed galaxies are demonstrably lacking in gas and devoid of dust, reddening and extinction. While this latter advantage holds true in the {\it application} of the TRGB method, the situation is somewhat more complicated when it comes to {\it calibrating} the TRGB. We consider this aspect of the TRGB calibration in what follows.

It is anticipated that the zero-point calibration of the TRGB will clearly be strengthened when absolute trigonometric parallaxes from Gaia become available for significant samples of nearby TRGB stars in the halo of the Milky Way. A preliminary calibration based on the Gaia Data Release 2 (DR2) has been published by Mould et al.~(2019), which is in broad agreement with earlier calibrations. However, the early-release parallax measurements are still preliminary; and, they are subject to known systematic uncertainties  (as discussed in Arenou et al. 2018). Anticipating the upcoming DR3 results from Gaia, which are still some years away from publication, we have, in the meantime, chosen to use the Large Magellanic Cloud (LMC) to set the absolute calibration for the TRGB. We have then compared this TRGB zero point with additional, parallel calibrations, each grounded in geometric distances (as is the LMC) derived from the analysis of detached eclipsing binaries (DEBs) in the SMC (Graczyk et al. 2014) and  47~Tucanae (Thompson et al. 2019). We have used the da Costa \& Armandroff (1990; hereafter DCA90) study of RGB stars in Galactic globular clusters, extending the analysis into the near-infrared using 2MASS data and using the new geometric distance to 47~Tucanae based on detached eclipsing binaries for the zero-point calibration.

For well-separated (detached) double-line spectroscopic, binary star systems, the stellar radii can be measured from their photometric and radial-velocity curves. As articulated by Paczyński (1997), the distance to the DEB can then be determined through an empirical calibration of the surface brightness (angular diameter) and color (temperature).  The DEB method has been recently and extensively developed by Pietrzyński and collaborators (see Pietrzyński et al. 2019 and references therein) concentrating on late-type giants. These authors find an extremely tight relation between the optical/near-IR $(V-K)$ color and surface brightness, with an $rms$ scatter of only 0.018~mag, from which they have determined a 1\% distance to the LMC.

Using the Hubble Space Telescope Advanced Camera for Surveys (\hstacs), Freedman et al. (2019) applied a calibration of the TRGB, which was subsequently applied to galaxies hosting Type~Ia supernovae (\sne), and used to determine a new value of the Hubble constant, anchored to the DEB distance to the LMC. In that work, we briefly described a new method for measuring the extinction for TRGB stars in the LMC, and we provided a preliminary calibration at $VIJHK$ wavelengths for the TRGB. Here we provide a more detailed description and update of the method, which makes use of a consistent set of TRGB stars with multi-wavelength data. {\it A clear advantage of this method is that it provides a direct measurement of the extinction, derived from, and applicable to the TRGB stars themselves.}

We first begin by illustrating  the expected behavior of the TRGB, based on published isochrones spanning a range in metallicity, color and age. Based on {\it VI} and {\it JHK} photometry for TRGB stars in the LMC we then detail our method for measuring the extinction and reddening, and determine a calibration of the TRGB zero point. We discuss our measurement of the LMC extinction in the context of other recent studies, provide two external consistency checks on our calibration, and present a value for H$_0$ based on our adopted calibration.

In Appendix \ref{app:appendix_Yuan}, we address some serious issues arising in the recent paper by Yuan et al. (2019; hereafter Y19). We demonstrate that the adoption of incorrect assumptions has led to an erroneous result; in specific, their underestimate of the line-of-sight extinction to the TRGB stars in the LMC.

\section{TRGB Theoretical Isochrones}

To provide some overall context for the method that we have developed, we first show {\it the expected behavior} of the TRGB as a function of metallicity, age and bandpass. For illustrative purposes, we use the PARSEC (Padova and Trieste Stellar Evolutionary Code: http://stev.oapd.inaf.it/cgi-bin/cmd)  isochrones (CMD Version~3.3; Bressan et al. 2012; Marigo et al. 2017). We are not aiming here to provide a comprehensive discussion of different TRGB models, but simply to convey, in broad outline, the general behavior of the TRGB as a function of wavelength, age and metallicity. We note that here and throughout this paper, all magnitudes are on the Vega System.

\begin{figure*} 
\centering 
\includegraphics[width=10.0cm,angle=-0]{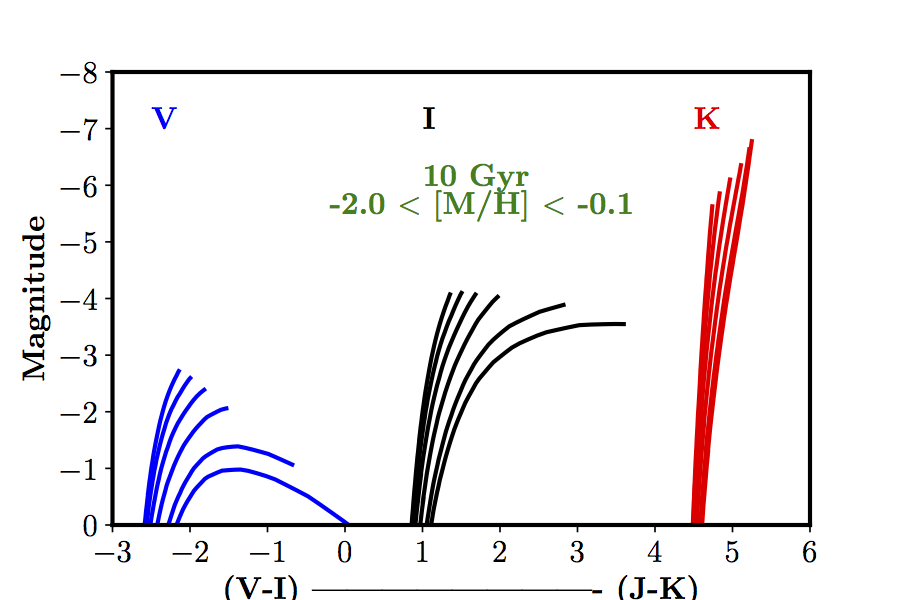}
\includegraphics[width=10.0cm,angle=-0]{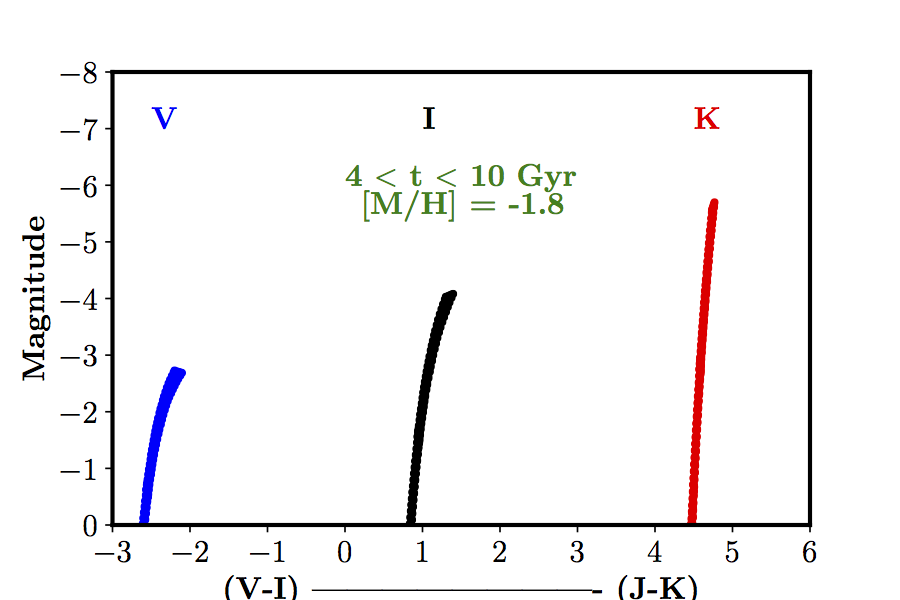}
\caption{\small  PARSEC isochrones for red giant branch stars with a constant age and a metallicity spread (upper panel) and a constant metallicity, but with an age spread (lower panel), are shown for  $V$ (blue), $I$ (black) and $K$ (red) bandpasses, to the same scale, for comparison. In the upper panel, the isochrones have a constant age of 10 Gyr and a metallicity range from -2.0 $<$ [M/H] $<$ -0.1 dex; in the lower panel, the isochrones have a fixed metallicity of [M/H] = -1.8, and an age spread of 4 $<$ t  $<$ 10 Gyr. The x-axis for the $V$ and $I$ isochrones is the $(V-I)$ color, while for $K$ it is $(J-K)$; however, for clarity the $V$-band isochrones have been shifted  by  -3.5~mag in $(V-I)$, and the $K$-band isochrones have been shifted by +4.0~mag in $(J-K)$. As can be seen, for older stellar populations ($>$4 Gyr), the RGB colors are very insensitive to age, while the colors track differences in metallicity very clearly. }
\label{fig:VIK}
\end{figure*}

In Figure \ref{fig:VIK} (upper panel), we examine the behavior with wavelength of red
giant branch stars with a {\it range of metallicities} at $V$, $I$, and $K$-band
wavelengths. Shown  are giant-branch isochrones with a large range of metallicities
spanning -2.0 $<$ [M/H] $< -$0.1 dex. As is well-known, with increasing metallicity the $V$-band isochrones become fainter with wavelength, the  $I$-band isochrones have a nearly-constant magnitude, and the
near-infrared isochrones become brighter. 

\begin{figure*} 
\centering 
\includegraphics[width=14.0cm, angle=-0]{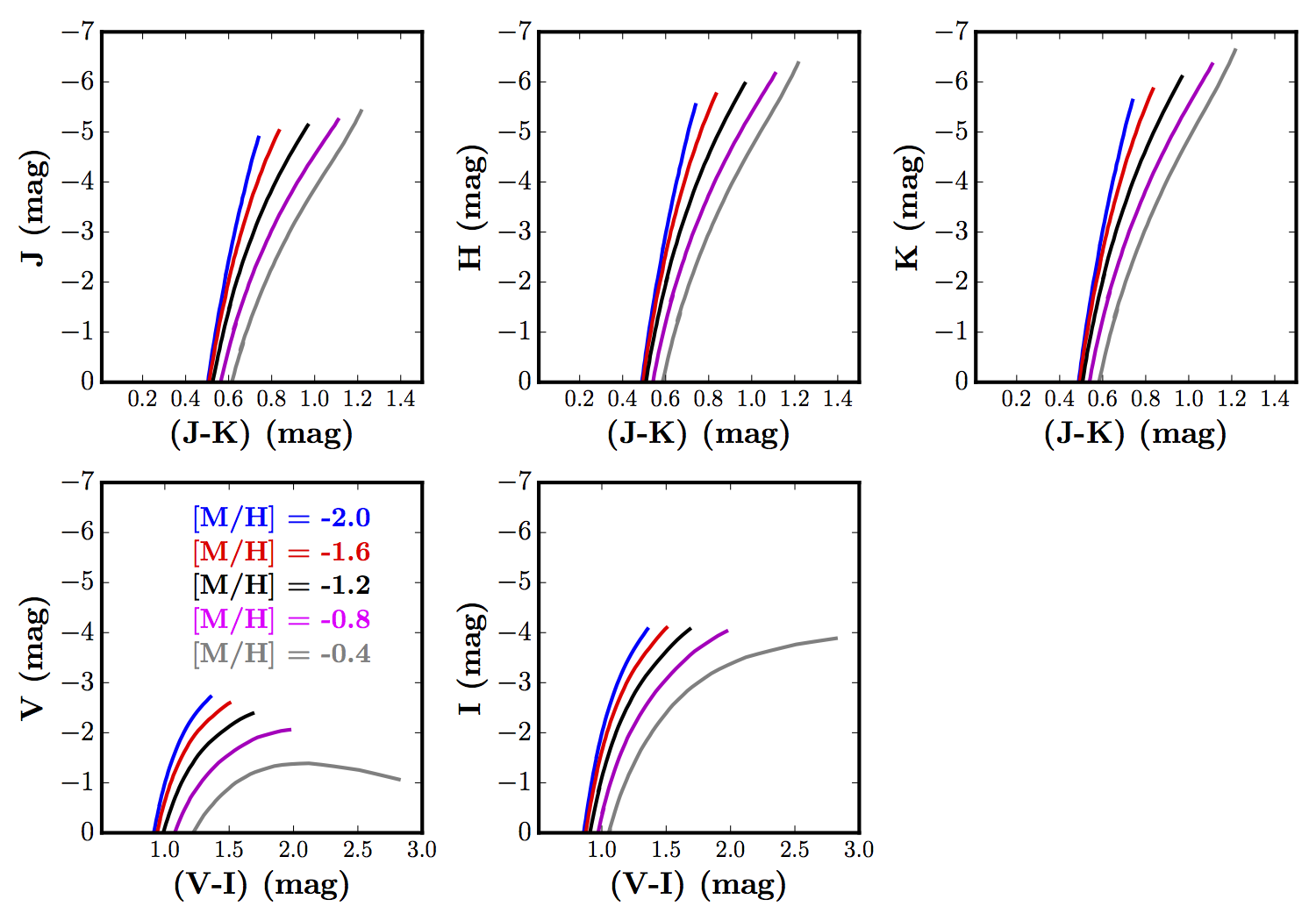}
\caption{\small PARSEC isochrones for a single 10 Gyr age at $VIJHK$ wavelengths. \bf{ The isochrones span a range of metallicities of  -2.0 $<$ [M/H] $< -$-0.4 dex.} These bandpasses correspond to those in the observed color-magnitude diagrams for the LMC, SMC and \ic shown in Figures \ref{fig:LMC_CMDs} and \ref{fig:SMC_1613_CMDs}. These isochrones show the well-known behavior of the TRGB stars as a function of wavelength: the tip stars increase in brightness in the infrared; they are nearly constant in the $I$-band, and they decrease in the $V$-band.}
\label{fig:VIJHK}
\end{figure*}

In Figure \ref{fig:VIK} (lower panel) we also examine the behavior with wavelength of red giant branch stars with a {\it range of ages}. Shown are giant-branch isochrones at $V$, $I$, and $K$-band wavelengths, with an age spread covering $4 \times 10^9$ to 10$^{10}$ years, at a fixed metallicity of [M/H] = -1.8 dex.   This plot illustrates the overall insensitivity of the giant branch luminosity, at a given wavelength, to the age of older RGB stars. Most of the color width of a composite old ($>$4 Gyr) stellar population results from a spread in metallicity, not a spread in age. At fixed metallicity an age spread of $4 \times 10^9$ to 10$^{10}$ years introduces an effective (color-correlated) magnitude scatter of $\pm$ 0.040  (at V), 0.005 (I), 0.012  (J), 0.023  (H) and 0.026 (K)~mag, respectively. In Figure \ref{fig:VIJHK} we show the isochrones varying in metallicity for a single 10 Gyr age at $VIJHK$ wavelengths. These wavelengths correspond to those for which we have data in the LMC, SMC and IC~1613, which we turn to in \S \ref{sec:method} below. 

We note here that the $I$-band magnitude is remarkably constant over this illustrated range of  metallicities (and corresponding colors), and gives rise to a single (standard candle)  $I$-band tip magnitude, while at other wavelengths there is instead a slanting distribution, or a (decreasing and/or increasing)``run" of -absolute magnitudes with increasing metallicities/colors. There is therefore no unique ``tip" magnitude in these other bands, but there is nevertheless a very well-defined (and theoretically predicted) correspondence of absolute magnitude with intrinsic color. We refer to these distributions as ``TRGB sequences". The same stars defining the tip of the red giant branch at a given metallicity will also define the tip  in different band color-magnitude diagrams, not at arbitrary magnitudes, but in the logical order that is predicted by the models. Moreover, the fact that higher metallicity manifests itself by making stars observed in the optical fainter while simultaneously resulting in brighter stars in the near-infrared, allows for the effects of metallicity and reddening to be decoupled and individually solved for. This direct dependence of the TRGB color with metallicity, and the negligible dependence on the age for older populations, provides a powerful means of accounting for differences in metallicity, while simultaneously correcting for extinction.

To summarize this section, there is little dependence on age for all TRGB stars older than about 4 Gyr (e.g., Serenelli et al. 2017). Conversely, there is an extremely tight relation  between the color of the stars defining the TRGB and the metallicity of the TRGB population, which provides a basis to empirically measure the extinction.

In what follows, we do not use the theoretical isochrones in our analysis to determine reddenings or distances: we undertake an entirely empirical analysis. However, we  confirm empirically that the behavior of the RGB with color and wavelength follows the general predictions from theory, thereby lending confidence to the overall method and its application.

\section{The Methodology: A Multi-Wavelength Approach to the TRGB Calibration}
\label{sec:method}

In this paper, we use $VI$ photometric data from the third phase of the {\it Optical Gravitational Lensing Experiment} (OGLE-III; Udalski et al. 2008) for both the LMC and the SMC. For the LMC, we use the ``Shallow" survey data published by Ulacyk, et al. (2012); both the LMC and SMC data are available at the following website http://www.astrouw.edu.pl/ogle/ogle3/maps/. Yuan et al. (2019) have pointed out that the coarser pixel scale in the optical ($UBVI$) survey of the two Magellanic Clouds undertaken by Zaritsky et al. (2002, 2004) probably suffer from blending issues. While the Zaritsky data for the SMC (but not the LMC) were used in the analysis by Freedman et al. (2019), here we now self-consistently use only the OGLE-III data for both the LMC and the SMC. In the near-infrared, $JHK$ observations of both Clouds were obtained in the course of the 2MASS all-sky survey. We have cross-identified the LMC and SMC stars from the OGLE-III surveys with those in 2MASS. This merged catalog forms the basis for the following study and the LMC calibration of the TRGB. The \ic data analyzed in this paper are from Hatt et al. (2017). 

The method that we are using is conceptually very simple. As noted above, the individual stars that define the tip in the $I$ band (at 8000\AA) are precisely the same stars that must also define the TRGB at both shorter and longer wavelengths.  Using a set of TRGB stars defined in the $I$ band, we then make use of those same stars at a range of wavelengths from the optical to the near-infrared, where the extinction decreases with increasing wavelength. The overall methodology is similar in concept (but not identical) to that developed in the past for determining extinctions to Cepheid variables (e.g., Freedman 1988, Freedman, Wilson \& Madore 1991).

Before proceeding further we note that in this study we confine ourselves to the color range 1.4 $<$ (V-I) $<$ 2.2~mag (-1.4 $>$ [Fe/H] $>$ -0.6), where the $I$-band magnitude of the TRGB is observed to be approximately constant (see F19 and references therein). At  colors beyond this range {\it the theoretical dependence} of the $I$-band TRGB is non-linear, as illustrated in Figure 3 of Mager, Madore \& Freedman (2008) (which, in turn, was derived from Bellazzini et al. 2001, 2008); however, the color range specified above encompasses virtually all of the low-metallicity halo stars used in extragalactic distance determinations, and it symmetrically straddles the peak of the Bellazzini relation in that color range. For all practical applications the TRGB $I$-band magnitude (in that restricted color range) is effectively constant; the formal scatter of less than $\pm$0.01~mag (deviating by only $\pm$0.015 mag, peak-to-peak).

\subsection{Schematic Illustration of the Method}
\label{sec:schematic_ext+red}

\begin{figure*} 
\centering 
\includegraphics[width=14.00cm,angle=-0]{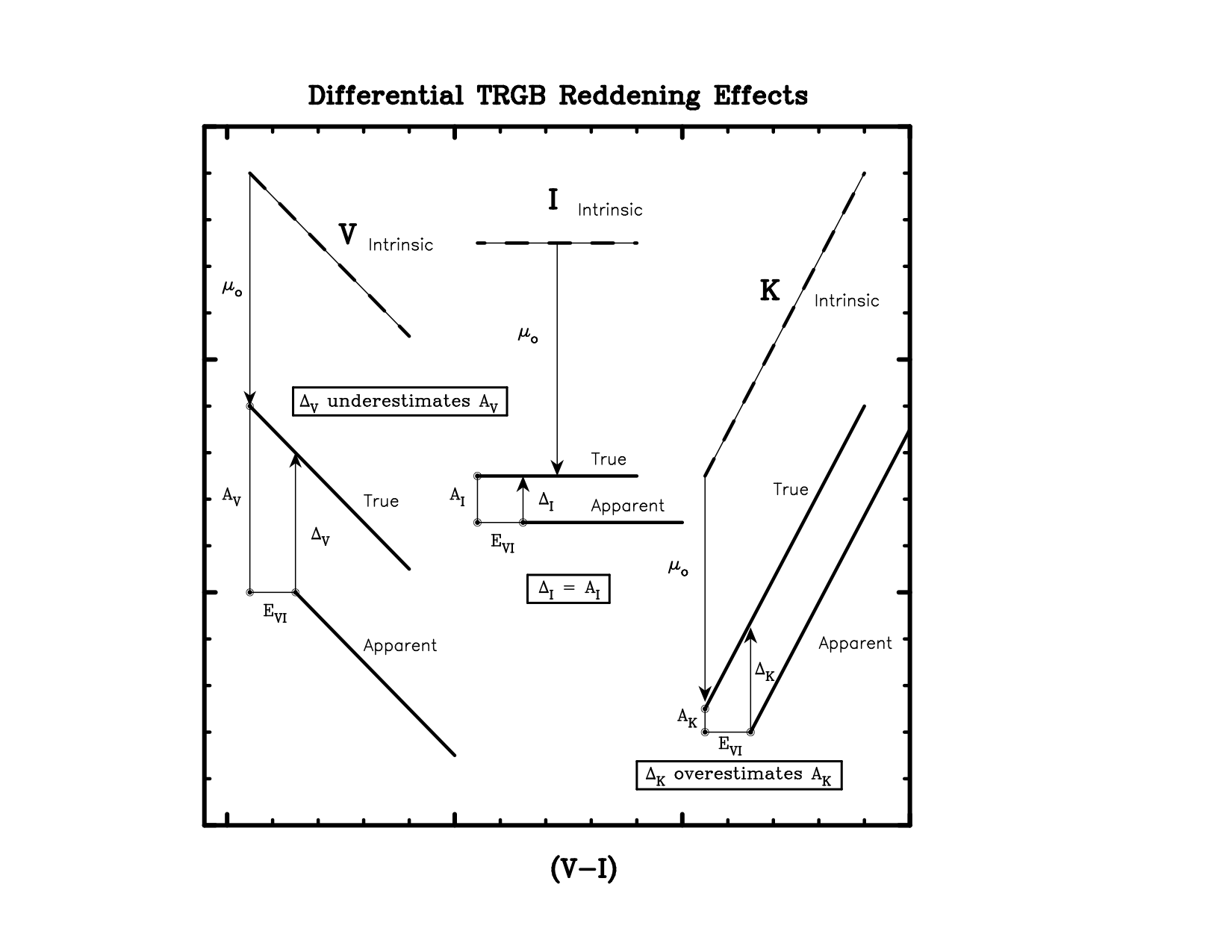}
\caption{\small A schematic illustration of multi-wavelength TRGB seen in three representative bands labeled V, I and K from left to right, illustrating the effects of distance, extinction and reddening. See text for details. }
\label{fig:schematic_ext+red}
\end{figure*}

In  Figure \ref{fig:schematic_ext+red}, we show a schematic representation of the steps involved in measuring both extinction and reddening. The upper dashed lines represent the run of the TRGB absolute magnitude as a function of intrinsic color. The $I$-band relation is flat with color. The brightest RGB stars decline in luminosity as a function of color for wavelengths shorter than the $I$ band, while these same stars increase in luminosity with color for wavelengths longer than the $I$ band. The solid lines (marked `True') midway down in each of the three diagrams are the intrinsic TRGB sequences displaced to fainter magnitudes by identical (true distance modulus) offsets labeled $\mu_o$. Further downward displacements of the TRGB lines are tagged by the wavelength-dependent extinctions ($A_V$, $A_I$ \& $A_K$) which systematically decrease in amplitude with wavelength, followed by wavelength-independent reddenings, each of the same amount, shown as $E(V-I)$ in each diagram. The resulting extincted, reddened and distance-modulus displaced traces of the three example TRGBs are labeled `Apparent'. The upward arrows separating the `Apparent' and `True' moduli in $V$, $I$ and $K$ have magnitudes of $\Delta_V$, $\Delta_I$ and $\Delta_K$ as marked. It should be noted that {\it these vertical displacements are not to be confused or directly equated with the extinction values appropriate to each of those wavelengths,  except for the $I$-band, where, coincidentally, the slope of the TRGB is flat with color}.  For all downward sloping (``blue'') TRGB relations $\Delta_{\lambda}$ underestimates $A_{\lambda}$; and for all upward sloping and (``infrared'') TRGB relations $\Delta_{\lambda}$ overestimates $A_{\lambda}$. The magnitude of that error (the difference between $A_{\lambda}$ and $\Delta_{\lambda}$)  is simply $s_{\lambda} \times E(V-I)$, where $s_{\lambda}$ is the slope of the TRGB in the selected color-magnitude diagram. For clarity, $\mu_{\lambda} = \mu_o +\Delta_{\lambda} - s_{\lambda} \times E(V-I)$.


\begin{figure*} 
\centering 
\includegraphics[width=14.00cm, angle=-0]{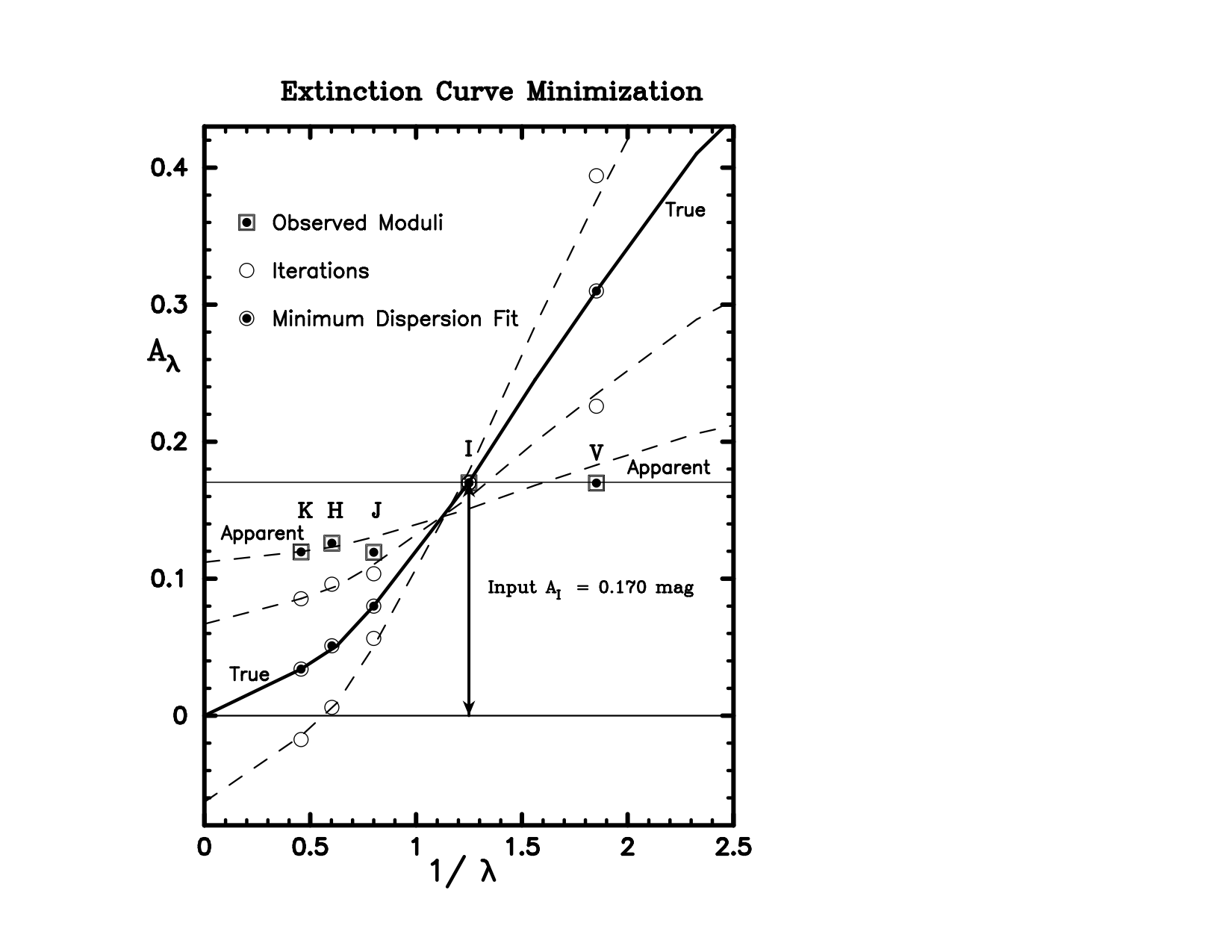}
\caption{\small A schematic illustration of the process of using TRGB data to 
simultaneously determine the true distance modulus and reddening from 
multi-wavelength data. 
Dots enclosed in squares are the result
of the first iteration on the reddening simply differencing the 
apparent magnitude of the tip stars with the intrinsic relations. 
The scatter is non-negligible.  Open circles 
show alternate reddening solutions on both sides of the input value. These solutions 
also show increasing (non-zero) scatter, most easily seen in the dashed-line solutions
passing above and 
below the $I$-band data point.
The input reddening is shown by the circled dots and fit by the steep solid line which 
was recovered by 
simply mimimizing the dispersion about the fit. See text for details.}
\label{fig:ext_sim}
\end{figure*}

Using simulated error-free data, Figure \ref{fig:ext_sim} illustrates the method for
using TRGB data to simultaneously determine the true distance modulus and simultaneously,
the total line-of-sight extinction and reddening from 
multi-wavelength data. The steep solid line (labeled True) is the run of extinction as a function of 
wavelength fit to the observations. Scatter around this particular fit is, 
by definition, zero for this simulated set of error-free data.  The apparent magnitude
off-sets $\Delta_{\lambda}$ 
between the intrinsic TRGB relations and the apparent TRGB relations (as described in
Figure \ref{fig:schematic_ext+red}) 
are shown as squares with inset filled circles. They show a shallow run of extinction with 
wavelength that is significantly less than the input value, and they show measurable scatter, resulting from not having applied the shifts required by the reddening terms. 
By iteratively compensating for the difference between $\Delta_{\lambda}$ and $A_{\lambda}$, 
re-plotting, re-fitting and each time re-calculating the sum of the squares of the residuals, a 
minimum is approached 
and then exceeded (when the applied reddening exceeds the input value), as illustrated by the 
two surrounding solutions in Figure \ref{fig:ext_sim} (which have measurable scatter clearly in excess of zero) shown as dashed lines through open circles.

\vfill\eject
\subsection{Application of the Method}
\label{sec:red_method}

\begin{figure*} 
\centering 
\includegraphics[width=14.0cm, angle=-0]{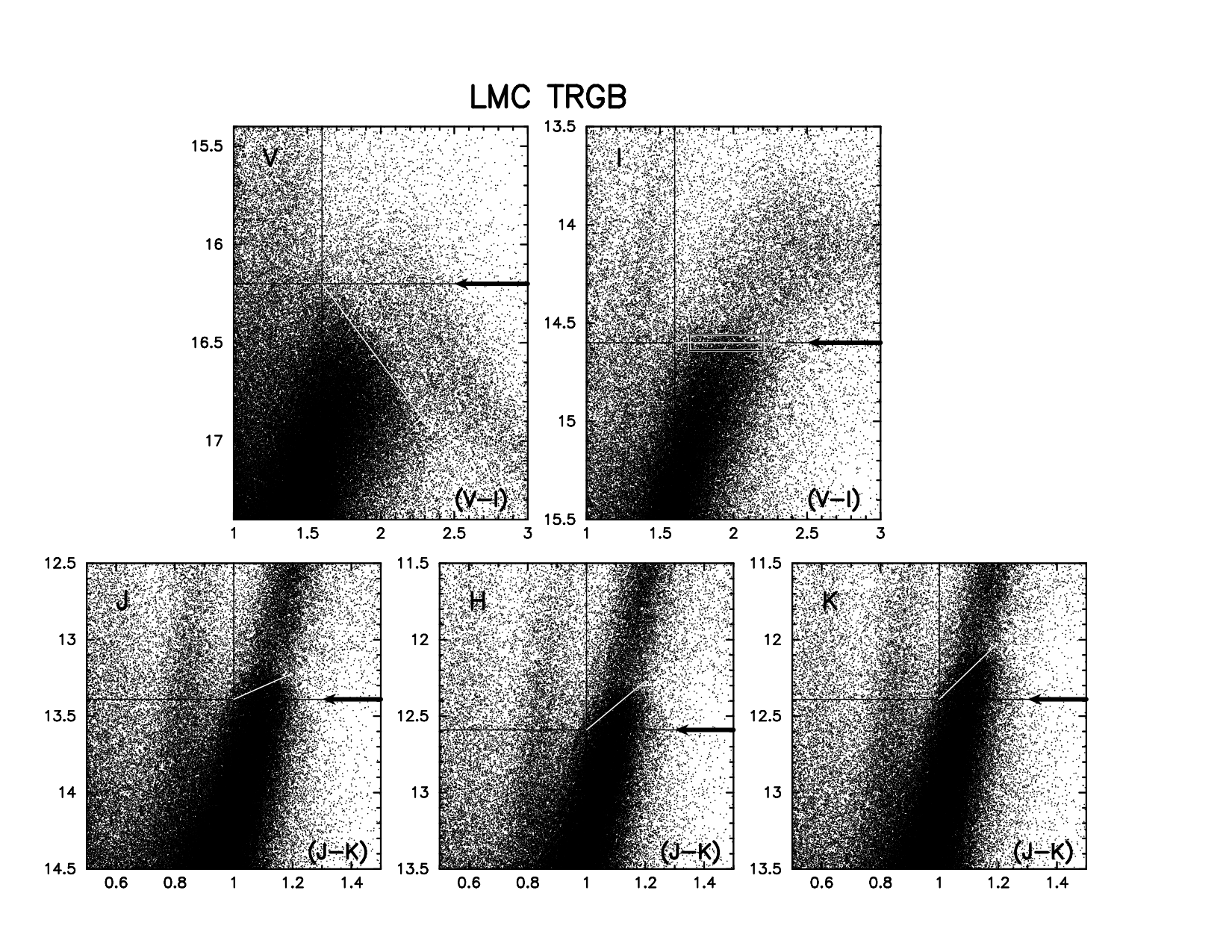}
\caption{\small Color-Magnitude Diagrams for the red giant 
branch population of stars outside of the bar region of the Large 
Magellanic Cloud. The two plots across the top are for the optical 
($VI$ band-passes and $(V-I)$ colors from OGLE-III), while the three across the bottom 
are for the near infrared ($JHK$  band-passes and $(J-K)$ color from 2MASS). All of the panels are 
zoomed into a two-magnitude vertical luminosity range and a lateral range in color 
of 2.0 mag in $(V-I)$ for the optical, and 1.0 mag in $(J-K)$ for the near infrared.
The white  lines mark the apparent magnitude level of the TRGB 
at each of the wavelengths, using predetermined slopes,  fit to the data
as described in the text. Arrows indicate the magnitude level in each 
bandpass at which the color calibration of the TRGB is normalized (i.e. at 
$(J-K)$ = 1.00~mag for $JH \& K$, and at $(V-I) = $ 1.80~mag in $V$ \& $I$.  The I-band selection function for the tracer stars is outlined by the white rectangle centrally-located in the upper right CMD.}
\label{fig:LMC_CMDs}
\end{figure*}

We begin with the  $I$ vs $(V-I)$ CMD where the slope of the TRGB is minimally impacted by color/metallicity variations, as noted above. The full set of CMDs for the OGLE-III LMC data is shown in Figure \ref{fig:LMC_CMDs}. For the LMC we spatially restricted the CMD sample to those stars {\it outside} of a one-degree radius circle centered on the bar of the LMC, where crowding within the high-surface-brightness (high-crowding) portions of the bar could bias the magnitudes of the TRGB stars therein.

\begin{figure*} 
\centering 
\includegraphics[width=10.0cm, angle=-0]{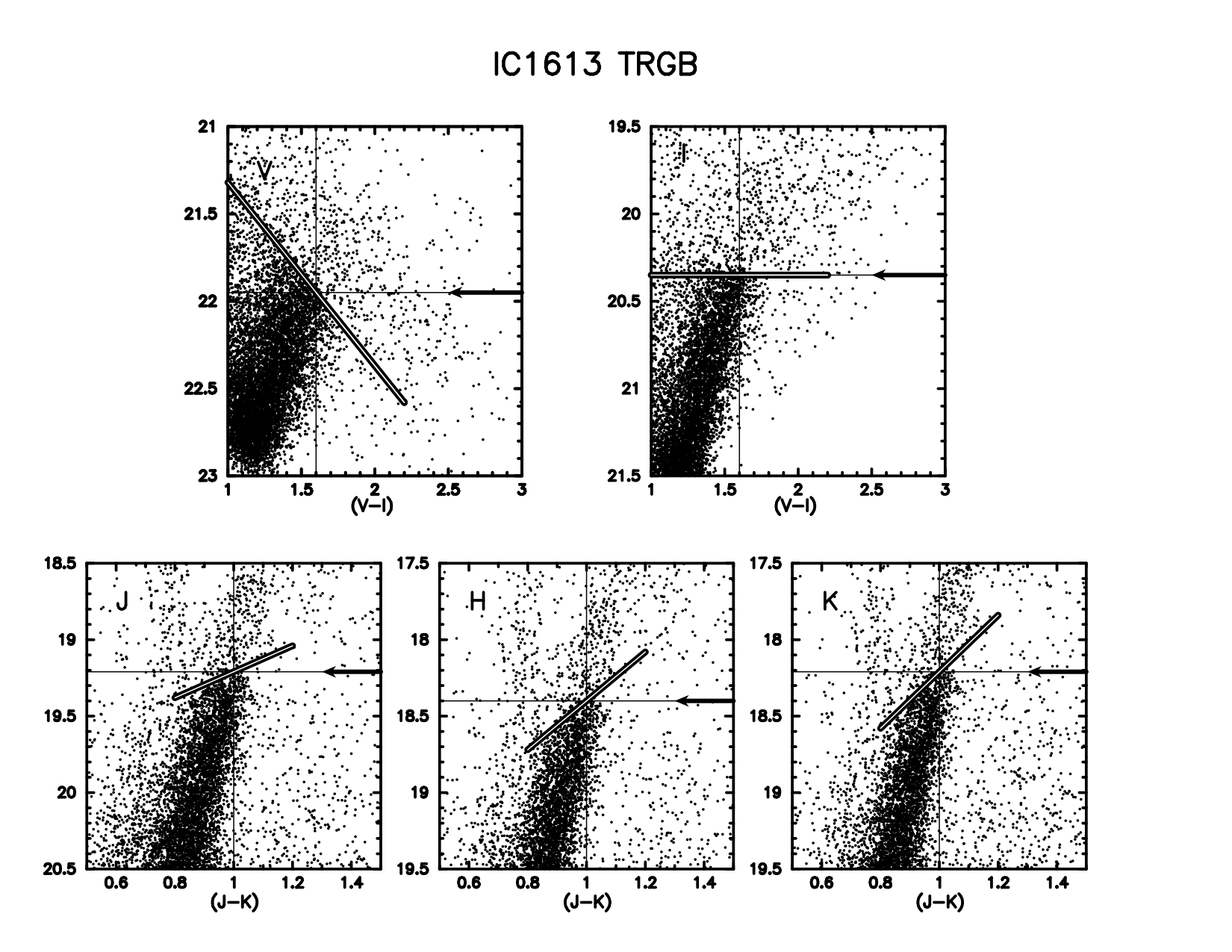}
\includegraphics[width=10.0cm, angle=-0]{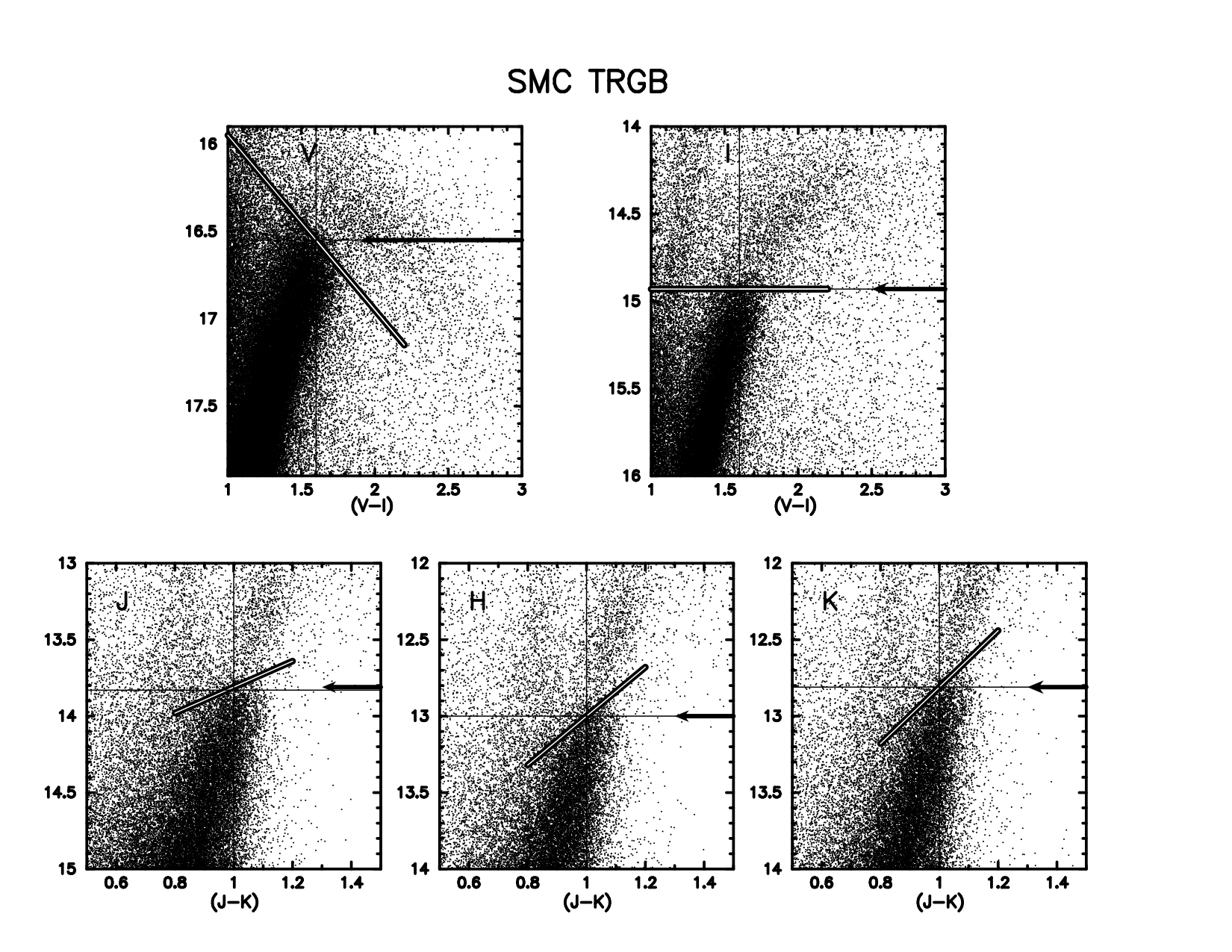}
\caption{\small Color-Magnitude Diagrams for the red giant branch population of stars 
in  IC~1613 (top five) and  the Small Magellanic Cloud (bottom five). The two across 
the top of each montage are for the optical ($VI$ band-passes, and ($V-I)$ colors), 
while the three across the bottom are for the near infrared ($JHK$  band-passes, and 
$(J-K)$ colors, from 2MASS). 
The IC 1613 data are from Hatt et al. 
(2017) and the SMC optical data are from OGLE-III. 
The slanting lines mark the apparent magnitude levels of the TRGB at each of 
the wavelengths, using predetermined slopes, and fit as described in the text. The arrows are as described in Figure \ref{fig:LMC_CMDs}. }
\label{fig:SMC_1613_CMDs}
\end{figure*}

The $I$ vs $(V-I)$ CMD is shown in the upper right panel. We selected a small 
subset of 306  tracer stars in the CMD that 
fall symmetrically about the calibrated fit to the tip, selected within the color range 1.4 $<$ (V-I) $<$ 2.2 mag. These stars have OGLE $I$-band photometry (as noted above, from the ``Shallow" survey data published by Ulacyk, et al. 2012), with quoted errors of less than $\pm$0.02~mag at I $\sim$ 14.5~mag. They sample the tip with a scatter of $\pm$0.04~mag, giving a scatter on the mean of $\pm$0.002~mag.  In the optical, the slope in the I-band has been set to zero (thereby resulting in a V-band slope of unity), while in the near-infrared the slopes are taken from Madore et al. (2018).  We give these here for completeness, along with their uncertainties: at $J, H$ and $K$, the slopes are -0.85 $\pm$ 0.09, -1.62 $\pm$ 0.16, and -1.85 $\pm$ 0.19, respectively. Our specific goal here is the difference in zero points, not a full solution of all possible parameters.  As we shall see, these slopes are consistent with the observed CMDs shown in Figures \ref{fig:LMC_CMDs} and \ref{fig:SMC_1613_CMDs} discussed below. Thus we solve here
for two parameters only: the total line-of-sight reddening and the true distance modulus for our selected 
sample of LMC TRGB stars.

Given the set of tracer stars described above (taken from Freedman et al. 2019 
and derived from their Figure 4), we mapped each of 
the $I$-band tracers into the CMDs at shorter ($V$) and longer
($JHK$) wavelengths. Using the adopted slopes of 
the TRGB in the flanking CMDs shown in Figure \ref{fig:LMC_CMDs}, 
we then determined their individual
zero points by minimizing the scatter between the intrinsic line 
of pre-determined slope and the tracer stars. Errors on the mean 
for each of the zero points were then calculated from the measured
scatter of the 306 tracer stars in each color. 

For our TRGB zero-point calibration via the LMC, we use two fiducial galaxies, IC~1613 and the SMC, each with very low line-of-sight reddenings. Before undertaking the multi-wavelength, differential reddening analysis of the TRGB population of the LMC with respect to these galaxies, we have first subtracted the small foreground reddenings of  $E(B-V)$ = 0.022 (\ic) mag and  0.033 mag (SMC),  respectively, based on the Galactic foreground reddening maps of  Schlafly \& Finkbeiner (2011), which is a re-calibration of Schlegel, Finkbeiner \& Davis 1998), as tabulated in NED.  To be clear, these maps do not include any correction for extinction in the main bodies of galaxies along the extended line of sight, but the host-galaxy TRGB stars are expected to have negligible extinction, given that they reside in the gas- and dust-free halos of these galaxies. In Figure \ref{fig:SMC_1613_CMDs}, we show a similar set of the five $VIJHK$ CMDs for IC~1613 and the SMC. The fits to their  TRGB sequences were obtained using the identical procedure as discussed above for the LMC stars. Subtracting the respective fits at each of the wavelengths for both galaxies, we determined differential apparent distance moduli to our subset of TRGB stars in the LMC for each of the five ($VIJHK$) bandpasses.  This step yields a first estimate of the differential distance moduli, each of which contains a fixed pillar, which is the difference in their true moduli. This common (geometric) offset is, of course, independent of wavelength. We emphasize, however, that these simple differential distance moduli at this stage do not yet account for the {\it reddening} of the TRGB stars, and they are therefore only a first approximation to the {\it extinction}. As we describe below, to measure {\it both extinction and total reddening} we iterate on the solution, solving simultaneously for distance, extinction and reddening by minimizing the  sum of the squares of the residuals around the fit of extinction versus inverse wavelength. In what follows, we have adopted the reddening curve used in NED which is taken from Schlafly \& Finkbeiner (2011), which in turn is derived from Fitzpatrick (1999) using $R_v = 3.1$. We have used a source spectrum appropriate to a K-type giant, closely matching  a typical elliptical galaxy, or as in this case, TRGB stars.

\vfill\eject
\subsection{Extinction Curves}
\begin{figure*} 
\centering 
\includegraphics[width=12.0cm, angle=-0]{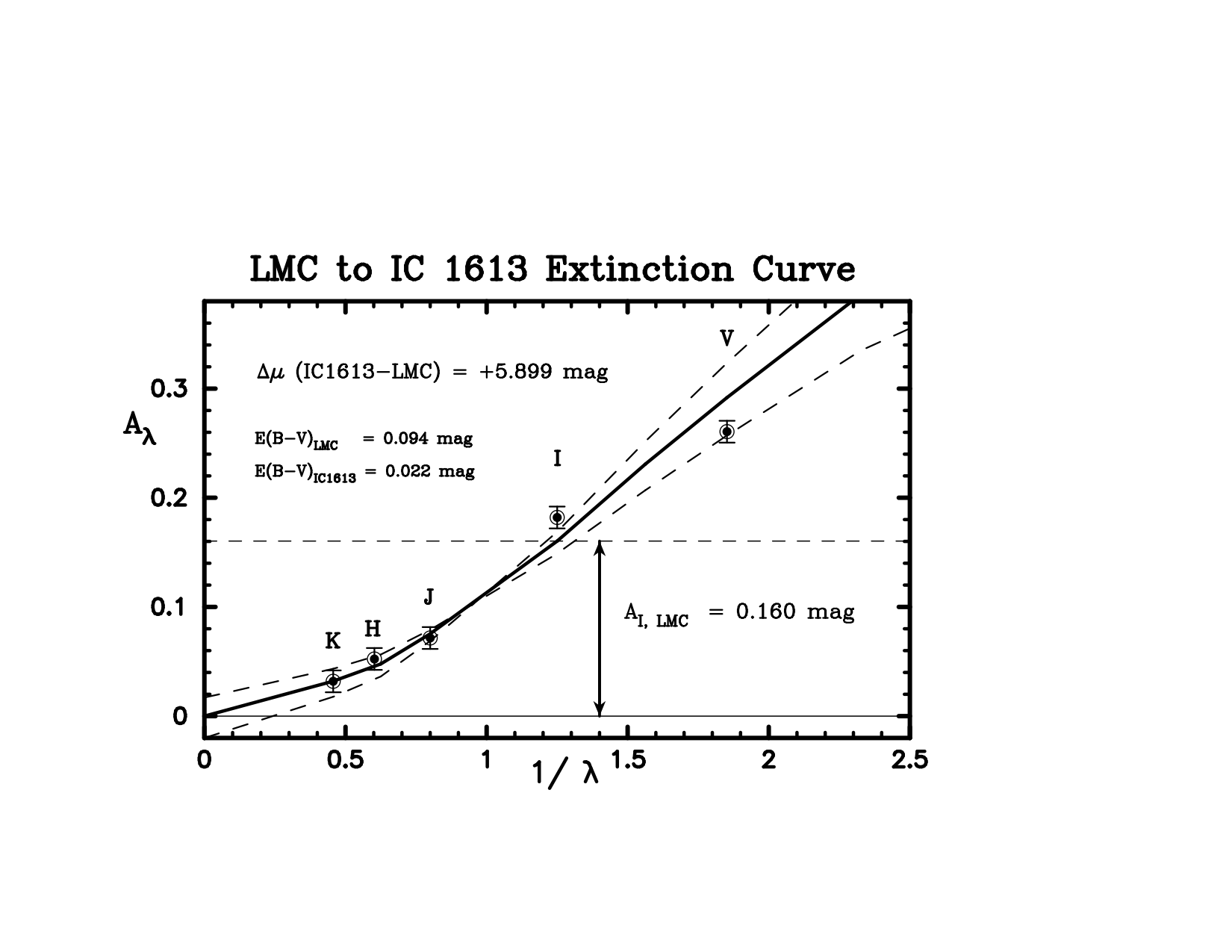}
\includegraphics[width=12.0cm, angle=-0]{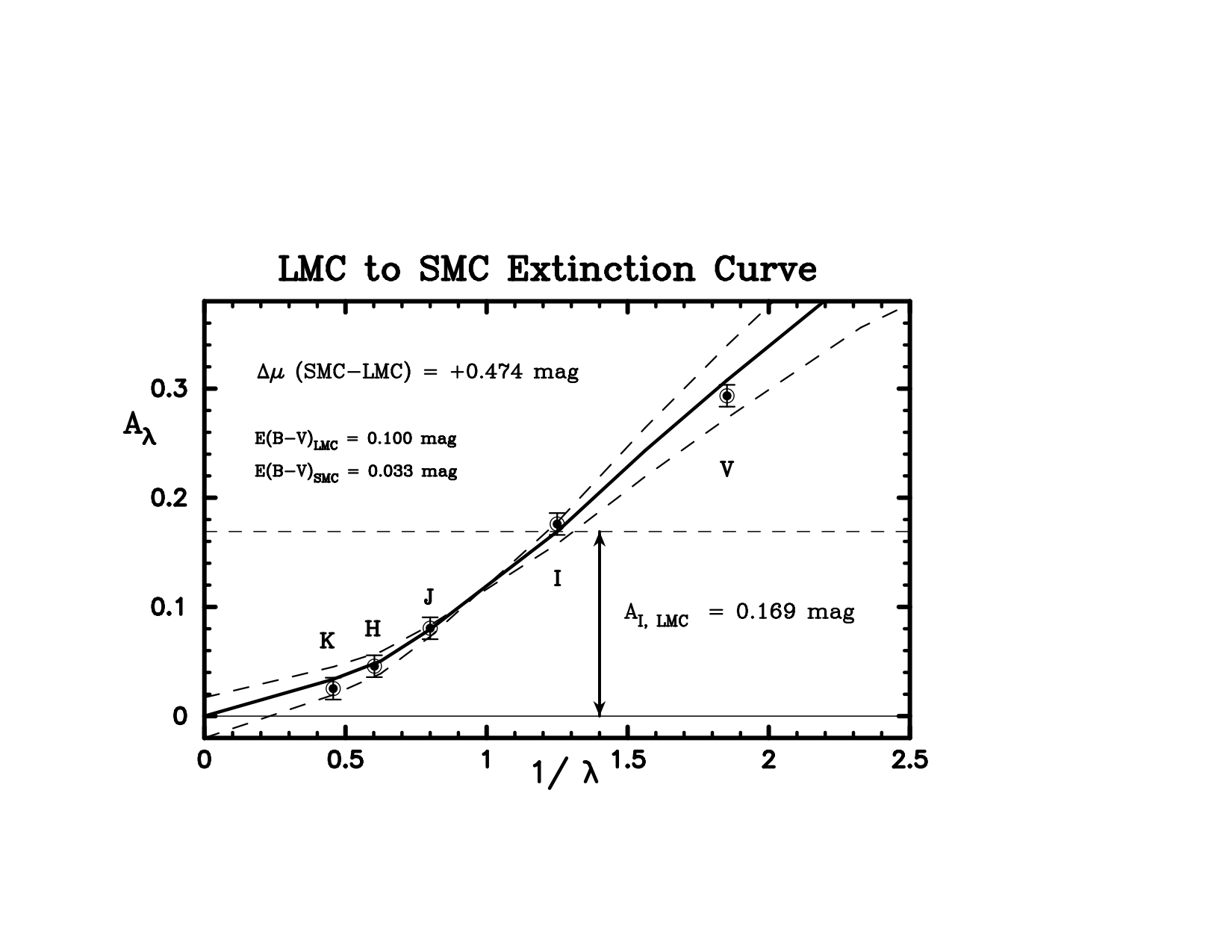}
\caption{\small The run of apparent differential distance modulus measured at five 
wavelengths ($VIJHK$) for LMC TRGB stars with respect to  IC~1613 and the SMC
in the upper and lower panels, respectively. 
The upward trending solid lines are Galactic extinction curves fit to the data 
where for the LMC $E(B-V) = $ 0.094~mag for the LMC adopting  $E(B-V) = $ 0.022~mag (from NED) for IC~1613 (upper panel); 
and $E(B-V) = $ 0.100~mag with $E(B-V) = $ 0.033~mag (from NED)  for the SMC  (lower panel), 
as described in the text.}
\label{fig:LMC_SMC}
\end{figure*}

In Figure \ref{fig:LMC_SMC} (lower panel)  
we show the iterated extinction plot derived from a comparison of the apparent magnitudes 
of the TRGB stars in the LMC with respect to \ic. The fit, as shown by the solid black line, 
gives a difference in their true distance moduli of  $\Delta\mu  =  $ +5.899~mag,
giving a true distance modulus for IC~1613 of $\mu_{IC1613} = $ 24.376~mag.
This distance agrees reasonably well with the recently-published 
Cepheid distance modulus of 
24.29 $\pm$ 0.03~mag given by Scowcroft et al. (2013).
(The mean and median of the 22 TRGB independently-determined values cited in 
NED are 24.340 and 24.365 mag, respectively.)
 The line-of-sight reddening to the LMC TRGB stars, using IC~1613 as the 
calibrator, gives for the LMC $E(B-V) = $ 0.094~mag ($A_I = $ 0.160~mag) which compares
favorably to the  value of $A_I = $ 0.169~mag derived below using the SMC. These values agree to within 0.001 to 0.002 mag with those given in Freedman et al. (2019); the small  difference results because a larger comparison sample of LMC stars (306 versus 200) was used in the current analysis.

Next we use the SMC as the normalizing source for a comparison with the LMC, which gives
an independent run of extinction with inverse wavelength for the LMC stars, as shown in
the lower panel of Figure \ref{fig:LMC_SMC}. 
Here we derive a difference in the true moduli of $\Delta \mu(SMC-LMC) = $ 0.474~mag
and a reddening to the TRGB stars in the main body of the LMC of E$(B-V)_{LMC} = $
0.100~mag  
($A_I = $ 0.169~mag). These values can be compared to the preliminary values 
in Freedman et al. (2019): $\Delta \mu(SMC-LMC) = $ 0.484~mag;  E$(B-V)_{LMC} = $
0.093~mag;  
($A_I = $ 0.158~mag). Here we adopt the NED foreground reddening to the SMC.   We note that the reddening value given in Freedman et al. (2019)
was E$(B-V)_{LMC}$ = 0.093~mag, and differs primarily as a result of our current adoption
of the OGLE-III data for the SMC rather than the photometry from Zaritsky et al. (2002). 

It should be noted that the two unknowns (true distance modulus and reddening)
cannot (mathematically) be determined from one pair of colors. At least three 
bands  (two independent colors) are required. If this minimal number is
adopted in future studies then, as large a wavelength baseline as possible 
should be chosen. With current facilities $V$, $I$ and $K$ bands would be a
competitive choice.

\vfill\eject
\subsection{Zero Points}
\label{sec:zero_points}

We adopt the DEB distance modulus to the LMC to be 18.477~mag (Pietrzyński et al. 2019). 
Taking the average of the two LMC reddenings derived above then gives $E(B-V)_{LMC} = $ 0.097~mag, 
which converts to $A_V = 0.300$, $A_I = 0.165$,   $A_J = 0.078$, $A_H = 0.049$ and $A_K = 0.032$
mag. Applying these corrections to the observed magnitudes of the TRGB stars in the LMC at their 
respective wavelengths gives the following absolute multi-wavelength calibrations of the TRGB 
method normalized to an intrinsic color of $(V-I)_o = $ 1.80~mag and applicable over the color 
range $ 1.4 < (V-I)_o < 2.2$ ~mag for the optical ($VI$), and normalized to $(J-K)_o = $ 1.00~mag 
for the near-infrared ($JHK$) calibration. In updating our calibration
with respect to earlier ones we have here rationalized the pivot-point colors in the optical and infrared so that the fits in both cases are set to colors that correspond to the same metallicity, i.e., [Fe/H] = -1.0 dex. This simply required moving the (V-I) normalization from 1.60 to 1.80~mag. This shift does not impact the $I$-band calibration, which is taken to be flat over the above color range. For the $I$-band, M$_I^{TRGB}$ = 14.595 - 18.477 - 0.165 = -4.047 $\pm$ 0.022 (stat) $\pm$ 0.039 (sys) ~mag, where the uncertainties are summarized in Table 2 of Freedman et al. (2019).

The difference in the current $(V-I)$ calibration from that given in Freedman et al. (2019) is due to  the small shift in M$_I$ from -4.04 (see Appendix A in Freedman et al.) to -4.05 mag, and re-centering on $(V-I)$ = 1.80~mag instead of 1.6~mag. The difference in the current $(J-K)$ calibration from that given in Freedman et al. (2019) is due to moving away from the LMC bar to a new outer annulus sample of stars, and ensuring that the M$_J$ and M$_K$ zero points differ by exactly 1.00 mag:

$$M_V = -2.25 + 1.00 \times [(V-I)_o - 1.8]$$
$$M_I = -4.05 + 0.00 \times [(V-I)_o - 1.8]$$
$$M_J = -5.14 - 0.85 \times [(J-K)_o - 1.0]$$
$$M_H = -5.95 - 1.62 \times [(J-K)_o - 1.0]$$
$$M_K = -6.14 - 1.85 \times [(J-K)_o - 1.0]$$

\subsection{Discussion of Published LMC Reddenings }
\label{sec:reddening}

\begin{figure*} 
\centering 
\includegraphics[width=16.00cm, angle=-0]{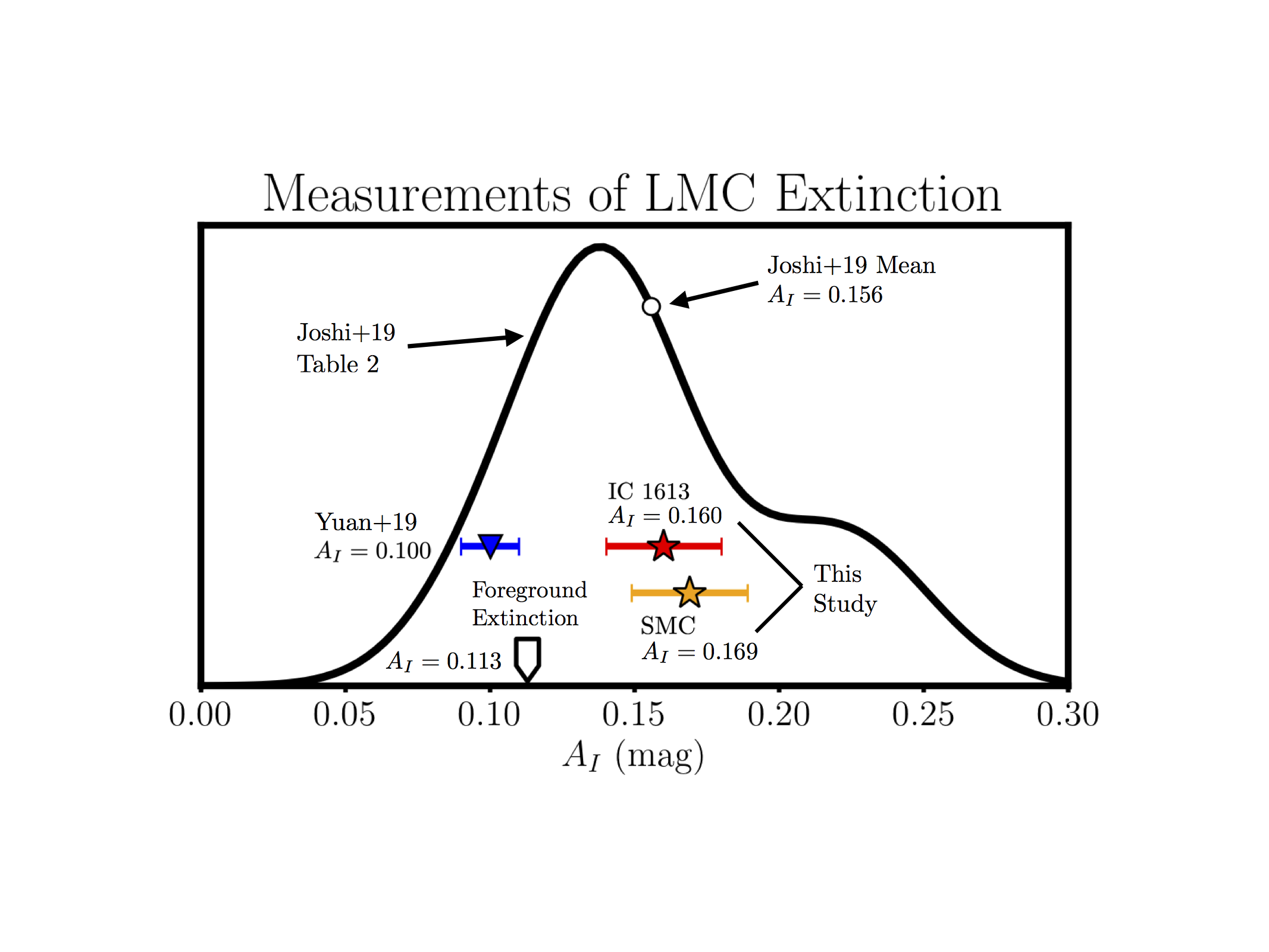}
\caption{\small  
Smoothed histogram of reddening values found in Joshi \& Panchal (2019), their Table 2,  converted to extinctions adopting a Fitzpatrick (1999) reddening law (solid black line). These extinction values are determined for a variety of stellar populations. The mean of the distribution is shown by an open white circle.
In blue is the extinction value suggested  by Yuan et al. (2019), based on a crowding correction, applied to the SMC photometry of Zaritsky et al. (2002). See Appendix \ref{app:appendix_Yuan} for a detailed description of why this post-processing of the results presented in Freedman et al. (2019) is problematic. In red and orange are the values determined here for TRGB stars, using  IC 1613 or the SMC, respectively, to set the reddening floor. These values agree at the millimag level with the equivalent measurements made in Freedman et al. (2019). For reference, the thick white arrow plotted is the Schlafly and Finkbeiner (2011) estimate of the foreground reddening to the LMC. We note that the {\it total} extinction adopted by Yuan et al. is smaller than  the foreground extinction. The two LMC extinctions derived in this paper, using the SMC and IC~1613, are shown as orange and red stars, respectively, and highlighted by two lines pointing to the label ``This Study".
}
\label{fig:AI_values}
\end{figure*}

Joshi \& Panchal (2019) have recently provided a summary of published spatial extinction maps for various populations of stars in the LMC (see their Table 2). In Figure \ref{fig:AI_values}, we show a smoothed representation of these published extinction values, which range from A$_I$ = 0.10 to 0.24 mag, after incorporating the detached eclipsing binaries at the midpoint of the range of values provided.  From a kernel density estimation, the mode of the $I$-band LMC extinction values is estimated to be A$_I$ = 0.138$^{+0.074}_{-0.029}$ ~mag (68\% CL). Our value lies close to both the mean and the mode of the distribution. We note that the range of values of  $A_I = $0.094-0.109~mag, chosen  by Yuan et al. (2019; hereafter Y19) for the LMC extinction, fall very near the lower bound on the extinction values quoted by Joshi \& Panchal, based on the Haschke et al. (2011) study. In their Section 4.1, Y19  quote an uncertainty  $\pm$0.03~mag (smaller than the uncertainty of 0.07 magnitude, quoted in the original Haschke et al. study), which renders this discrepancy even more significant. 
Finally, we further point out that the Y19 adopted value falls below the average foreground extinction of $A_I =$~0.113~mag along the line of sight to the LMC (Schlafly \& Finkbeiner 2011).

\section{Consistency Checks on the TRGB Calibration}
\subsection{DEB Distance to the Small Magellanic Cloud (SMC)}

Graczyk et al. (2014) used five DEBs in the SMC to determine the differential distance between the two Clouds deriving a value of  $\Delta \mu(SMC-LMC) = $ 0.472~mag $\pm$ 0.025~mag (lower limit based on the SMC uncertainty alone) using an older DEB distance to the LMC. Each of the individual DEB measurements have a typical quoted uncertainty of $\pm$ 0.03 mag. Updating to the most recent LMC DEB distance scale of $\mu_o = $ 18.477 $\pm$ 0.024~mag (Pietrzyński et al. 2019), this gives  $\Delta \mu(SMC-LMC) = $ 0.488 $\pm$ 0.035~mag, which is within half a sigma of our  independent determination of 0.474~mag, as given above. (See Pietrzyński et al. (2019) for a recent and detailed discussion of the DEB method as applied to the calibration of the extragalactic distance scale.) Updating the Y19 calculation in their Section 4.2 increases the extinction to the LMC to $A_I$ = 0.125~mag, and places it beyond the 3$\sigma$ contours of their Figure 10 solutions.

Adopting $\Delta \mu(SMC-LMC) = $ 0.488 ~mag and the DEB distance modulus for the SMC of 18.965 $\pm$ 0.025 (stat) $\pm$ 0.048 (sys)~mag, our detection of the I-band TRGB at 14.93~mag and an extinction of $A_I = $0.056~mag (NED) gives $M_I(SMC) =$ -4.09 $\pm$ 0.03 (stat) $\pm$ 0.05 (sys)~mag, in systematic agreement, at the one-sigma level, with our LMC measured value of -4.05 $\pm$ 0.03 (stat) $\pm$ 0.05 (sys)~mag. The SMC tip detection was made using a standard Sobel filter applied to  a GLOESS-smoothed I-band luminosity function based on stars within the color range 1.4 $< (V-I) <$ 1.8~mag, using software and methods identical to those used in Freedman et al. (2019). Within that same color range 138 tip tracer stars, used to measure the tip magnitudes in VJH \& K, yielded an error on the mean of $\sigma_{I(TRGB)}$ = ~$\pm~$0.007~mag.

\vfill\eject
\subsection{DEB-Based Galactic Globular Cluster Calibration of the TRGB}

We now present a second independent check on the calibration of 
the TRGB. The zero point of this calibration of the near-infrared TRGB is based upon the recently measured detached eclipsing binary (DEB) based geometric distance to 
47~Tucanae (Thompson et al. 2019).\footnote{We note that, in general, the limitation of measuring the TRGB in individual globular clusters is the small numbers of stars populating the tip. These measurements can, however, serve to provide a firm lower limit to the tip magnitude. However, in the more populated clusters (e.g., 47 Tucanae and $\omega$ Cen), the numbers are large enough to provide calibrations in their own right.} 

Following da Costa \& Armandroff (1990; DCA90) we use a selection of Galactic globular clusters, covering a wide range of metallicities ([Fe/H] = -2.2 to -0.7)\footnote{As compiled by Harris (2010): http://physwww.mcmaster.ca/~harris/mwgc.dat} to produce
multi-wavelength ($JHK$) composite color-magnitude diagrams.   The composite CMD can then serve as the basis for an independent check on the zero-point calibration of the TRGB sequences, for comparison with the mixed-metallicity populations seen in the halos of nearby galaxies. Our sample includes M2, M4, M5, M15, M55,  NGC~0362, NGC~1851, NGC~6362, NGC~6397,  NGC~6752, and 47~Tuc. We use the DCA90 compilation because it provides a set of consistent {\it relative} distances to the various clusters. We set the absolute zero point using the recently determined DEB measurements. In a more detailed, multi-wavelength study of the TRGB calibration in preparation, we will be applying proper-motion selection using high-resolution Gaia DR2 data to study a significantly larger sample of 45 Galactic globular clusters (observed in 10 band-passes from the optical to the near-infrared).

At present, we have revisited the TRGB calibration of the slopes and zero points primarily in the near-infrared for two main reasons: (1) There is homogeneous, high precision and high accuracy $JHK$ photometry for all of these targets, obtained in the course of the 2MASS Survey all-sky (Cutri et al. 2006).\footnote{These data are publicly available through $IRSA$ at https://irsa.ipac.caltech.edu/cgi-bin/Gator/nph-dd.} 
(2) The impact of line-of-sight reddening on the zero-point 
calibration of the TRGB is greatly diminished by working in the infrared. The globular cluster sample studied by DCA90 is a relatively low 
line-of-sight reddening subset of Galactic globular clusters, with reddenings ranging from $E(B-V) = $0.02 to 0.10~mag, which translate to extinction corrections in the near infrared of  $A_J = $ 0.016 to 0.080~mag, $A_H = $ 0.01 to 0.05~mag and $A_K = $ 0.007 to 0.034~mag (after dropping NGC~6397 because of its large foreground extinction 
of $A_V = $ 0.56~mag.) Reddening corrections have been adopted from DCA90. 
We note also the {\it JHK} TRGB study for 24 globular clusters belonging both to the bulge and the halo of the Galaxy,  undertaken by Valenti et al. (2004a,b). In general the reddenings for this sample are larger (with $E(B-V)$ values up to 1.3~mag), so we have not (at this time) used them for our purposes.
\vfill\eject
\subsubsection{47 Tucanae}

Thompson et al. (2019) have determined direct distances to two DEBs in the metal-rich Galactic globular cluster 47~Tuc. These two stars have geometrically determined distances  of 4.41 $\pm$ 0.06 and 4.60 $\pm$ 0.09 kpc, respectively, giving an average true distance modulus of $\mu_o =  13.27 \pm$~0.07~mag. As noted in Thompson et al. the analysis by Chen et al. (2018) of Gaia parallaxes to 47~Tuc yield a distance modulus of 13.24 $\pm$ 0.005 (stat) $\pm$ 0.058~mag (sys). The Gaia parallax corresponds to a distance of 4.45~kpc, which agrees with the DEB distance to within 1\%. Adopting the DEB geometric calibration, we re-zero the true distance moduli to each of the DCA90 globular clusters, with the required offset being -0.12~mag.

\subsubsection{Composite Globular Cluster Color-Magnitude Diagrams and Luminosity Functions}

\begin{figure*} 
\centering 
\includegraphics[width=8.0cm,angle=-0]{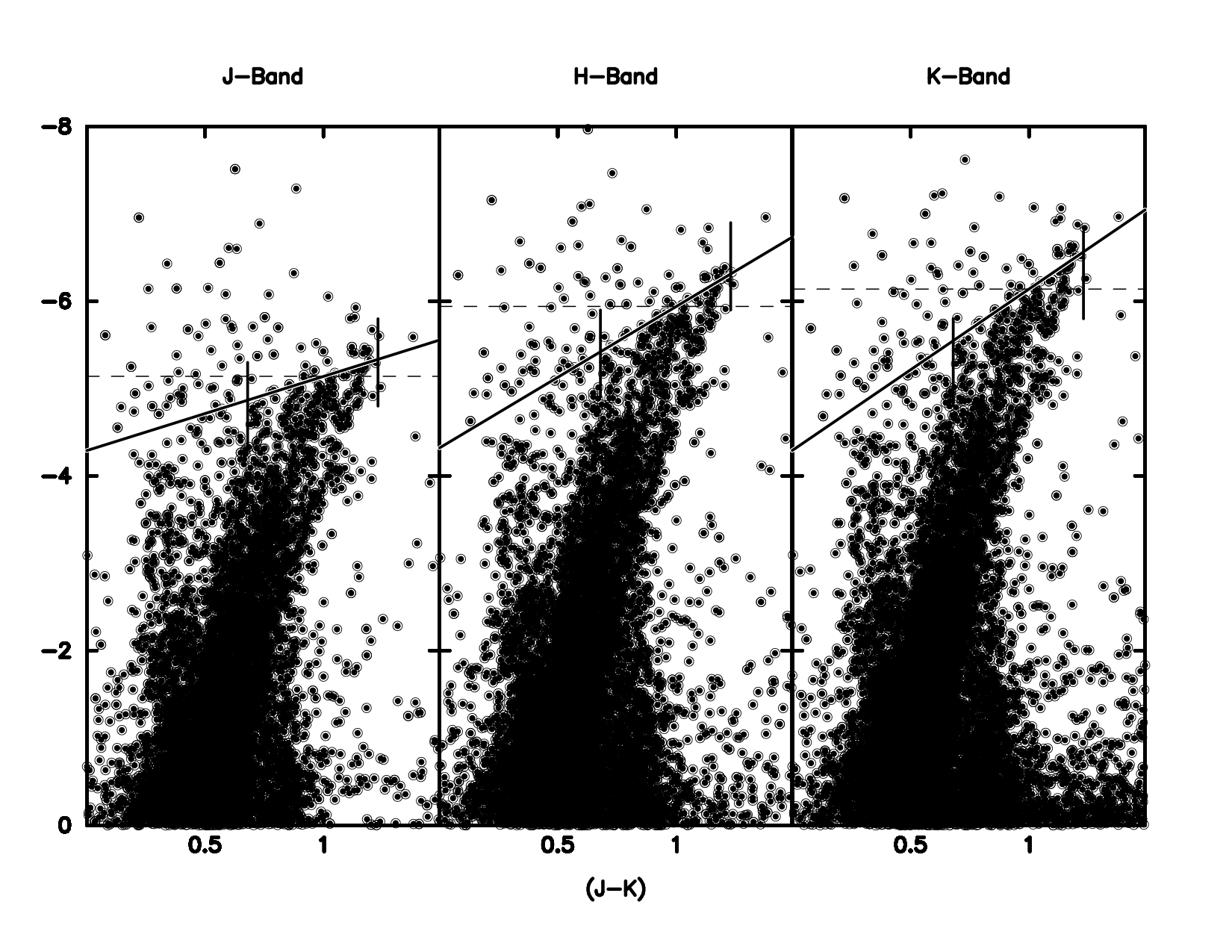}
\caption{A composite CMD for the 11 Galactic clusters.  The upward-slanting lines are fits to the TRGB sequences in Madore et al. (2018), as described and updated in the text. The dashed line is the fiducial magnitude read off at $(J-K)_o = $ 1.00~mag. The vertical lines at $(J-K) = $ 0.7 and 1.2~mag mark the color range within which the linear approximation to the TRGB sequence holds observationally.}
\label{fig:comp_JHK}
\end{figure*}

In Figure \ref{fig:comp_JHK} we show a composite $JHK$ CMD which combines each of 
the individual datasets for the least-reddened clusters. 
The upward-slanting lines in each of the three panels trace the TRGB 
calibration, previously published 
in Madore et al. (2018). The correspondence between the two provides an 
independent check and confirmation of the result derived from 
the LMC TRGB stars. The two vertical lines mark the intrinsic color range 
(0.65 $< (J-K) <$ 1.25 mag) over which a linear calibration can be  
observationally defined. The horizontal broken line shows the magnitude 
of the TRGB increasing in brightness towards longer wavelengths 
as measured at the fiducial color of $(J-K) = $ 1.00~mag.

\begin{figure*} 
\centering 
\includegraphics[width=8.0cm,angle=-0]{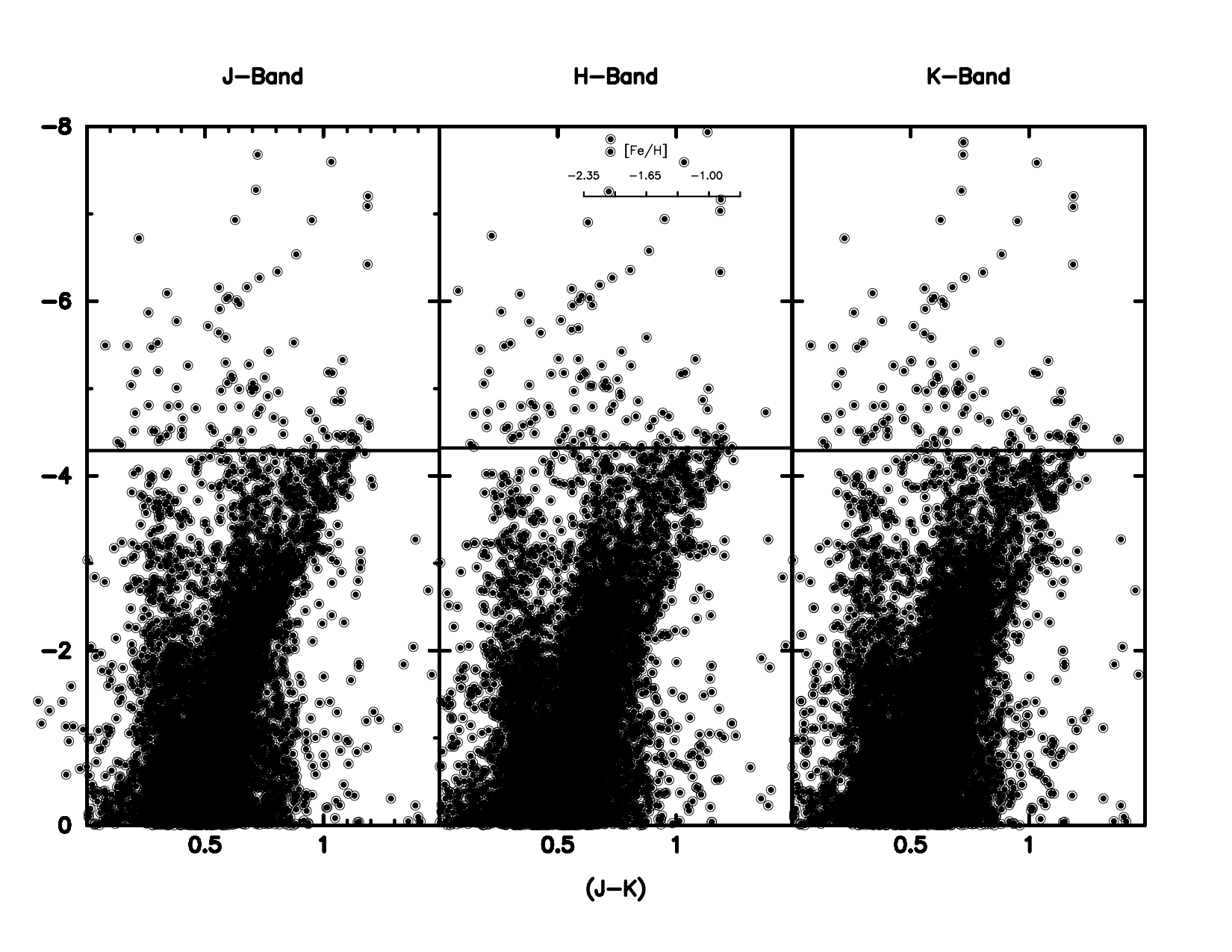} 
\caption{A composite CMD for the clusters  shown in Figure \ref{fig:comp_JHK} , where in this case,  the magnitudes have been rectified by flattening out the color dependence seen in the raw CMDs. }
\label{fig:comp_rect}
\end{figure*}

\begin{figure*} 
\centering 
\includegraphics[width=8.0cm,angle=-0]{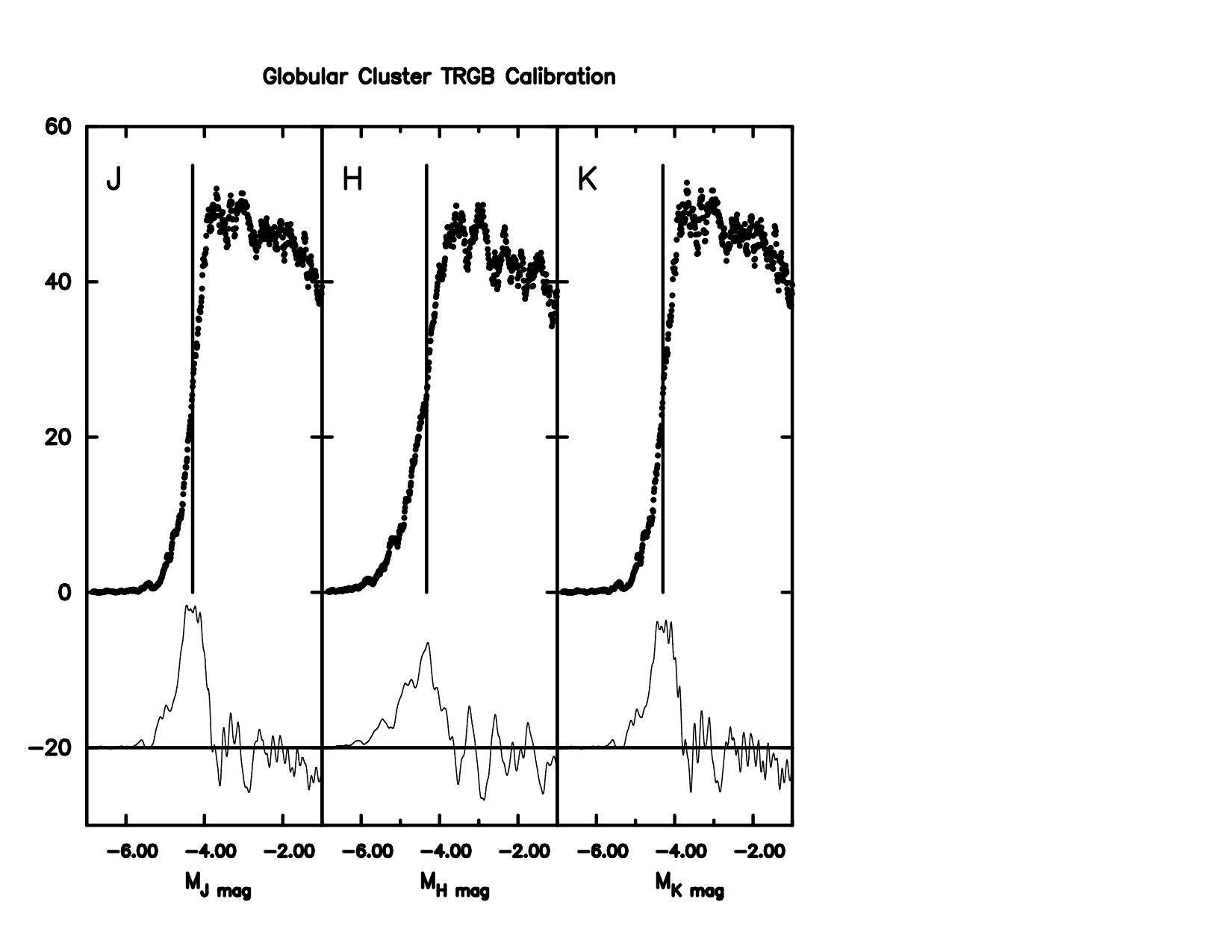} 
\caption{Galactic Globular Cluster Edge-Detector Output. $JHK$ luminosity functions for the rectified CMDs (using the Madore et al. 2018 slopes) (upper curves) and their corresponding edge-detector responses (lower curves) are given for a six-magnitude interval in luminosity, centered on the TRGB. See text for more details.}
\label{fig:comp_edge}
\end{figure*}

Using the measured slopes we flatten or ``rectify'' the color-magnitude 
diagrams such that their marginalized luminosity functions show the greatest
contrast in the TRGB discontinuity. The rectified CMDs are shown in Figure
\ref{fig:comp_rect}. 
Figure \ref{fig:comp_edge} shows the marginalized luminosity functions of the
rectified composite CMDs for the Galactic globular clusters shown above. 
117 RGB stars contribute to the $\pm$0.3 mag interval within which the detection
and measurement of the tip was made in this figure. The statistical uncertainty
on the detections is found to be $\pm$0.051~mag in $J$ and $K$
and $\pm$0.057~mag in $H$. Averaged over the three wavebands, the derived zero
point in the $I$-band is calculated to be $M_I = $ -4.056 $\pm$ 0.053 (stat)
$\pm$ 0.080 (sys) mag, where the systematic error is that carried over directly from the DEB systematic error on the distance to 47~Tuc.

\subsection{Geometry of the LMC}

A further test that we have undertaken is to determine the effect of  geometry  on the measurement of the TRGB in the LMC. Using the OGLE III photometric data from Ulaczyk et al. (2012), we adopted a position angle of 132 deg and an inclination angle of 25 deg for the LMC disk, as determined by Pietrzyński et al. (2019)  (see also Figure 3 of Jacyszyn-Dobrzeniecka et al. 2016,    based on Cepheids). As these and earlier studies of the LMC have shown, the northeast side of the LMC is closer to us.

\begin{figure*} 
\centering 
\includegraphics[width=14.0cm,angle=-0]{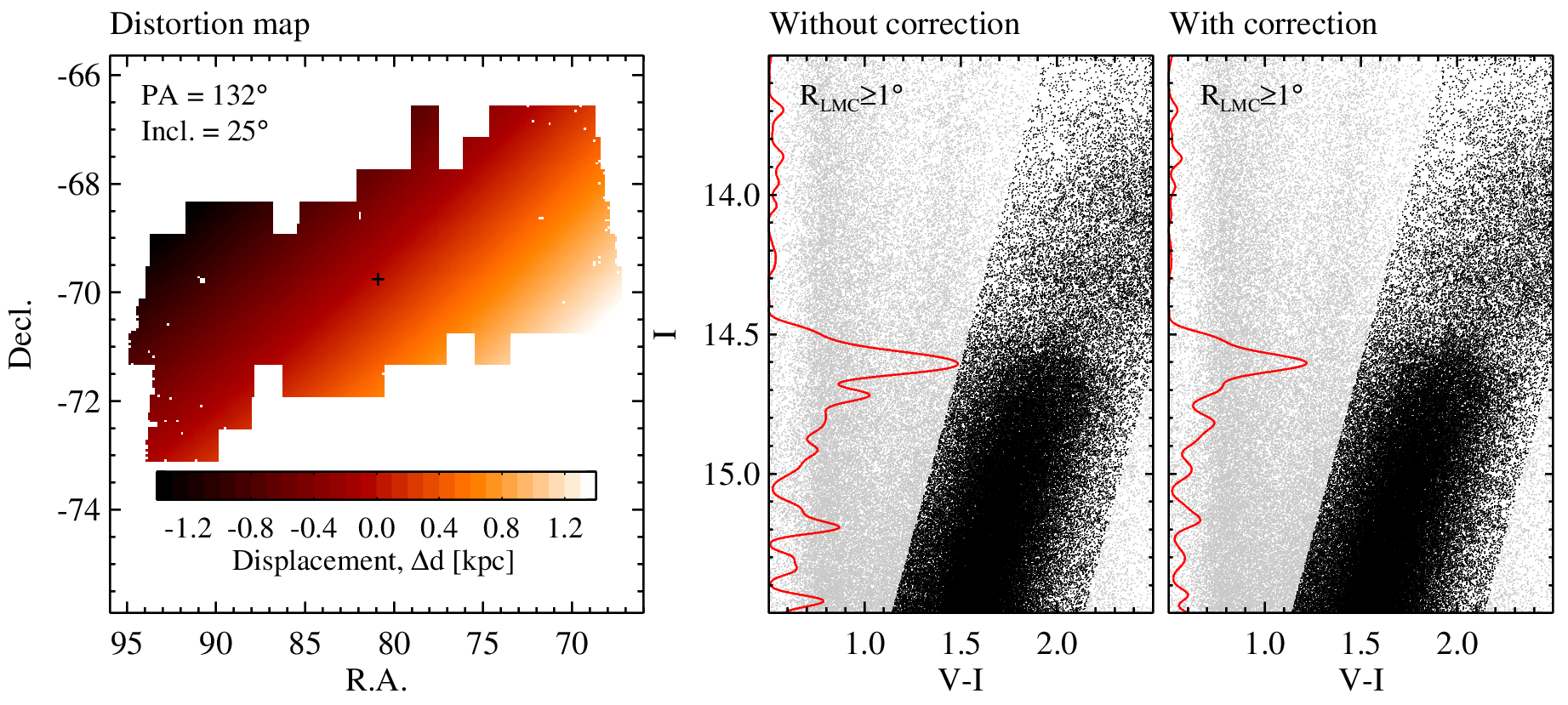} 
\caption{Test of the effects of geometry on measurement of the TRGB in the LMC. (Left) a map showing the displacement ($\Delta d = d - d_{corr}$) along the line of sight direction, as a result of the tilted LMC disk. We assumed a position angle of 32 deg and an inclination angle of 25 deg (Pietrzyński et al. 2019) for the LMC disk plane. The mean of the displacement is only $\sim$10 pc. (Right two panels) CMDs of the outer region of the LMC before (left) and after (right) the distortion correction. The measured TRGBs are almost identical (14.606 mag and 14.604 mag), respectively, indicating  that the LMC disk geometry has no statistically significant impact on our TRGB measurement. 
}

\label{fig:LMCtilt}
\end{figure*}

We calculated the displacements of individual stars along the line of sight direction between the tilted (observed) and non-tilted (inclination corrected) disks. The displacement is measured to be up to $\sim$1400 pc ($\sim$3\% in distance and $\sim$0.06 mag in distance modulus). The median/mean of the displacements is very small, amounting to only $\sim$10 pc. This is naturally expected, as the spatial coverage of the OGLE data is quite symmetric centered on the LMC. We then applied the edge detection algorithm to the original and the inclination-corrected OGLE catalogs. As can be seen in Figure \ref{fig:LMCtilt}, the two measured tip values are entirely consistent, with values of 14.606 mag and 14.604 mag, respectively,  with uncertainties of 0.02~mag.

\vfill\eject
\subsection{Summary of Consistency Checks on the Zero Point}

As summarized in Table \ref{tab:zeropoints}, the independent determinations of the $I$-band magnitude for the TRGB based on the 1) DEB measurements for the SMC and 2) {\it JHK} 2MASS observations for a sample of Galactic globular clusters, calibrated by DEB distance measurements in 47 Tucanae, agree very well with our determination for the LMC (see \S\ref{sec:zero_points}). We adopt the latter measurement, M$_I^{TRGB}$ = -4.047 $\pm$ 0.022 (stat) $\pm$ 0.039 (sys) for our calibration of the TRGB. For clarity, we note that the calibration for the F814W filter, differs from the $I$-band calibration by -0.0068 mag (see Freedman et al. 2019, \S3.4), giving M$_{F814W}^{TRGB}$ = -4.054 $\pm$ 0.022 (stat) $\pm$ 0.039 (sys).

\begin{deluxetable}{lcccl} 
\tabletypesize{\normalsize}
\setlength{\tabcolsep}{0.05in}
\setlength{\tabcolsep}{12pt} 
\tablecaption{TRGB Zero Points \label{tab:zeropoints}}
\tablewidth{0pt}
\tablehead{ \colhead{Data}  &  \colhead{M$_I^{TRGB}$} & \colhead{$\sigma_{stat}$} & \colhead{$\sigma_{sys}$}& \colhead{Notes}  }
\startdata
        LMC    &  -4.047 &	    0.022  &	    0.039 & Comparisons with \ic and SMC\\
        SMC   &  -4.09  &     0.03    &	    0.05 &  SMC DEBs \\        
        Globular Clusters   &  -4.05  &     0.05  &	    0.08   &  47~Tuc DEBs + Composite {\it JHK}\\ 
\hline \hline
\enddata
\end{deluxetable}

\section{Implications for the Hubble Constant}
\label{sec:ho}

\begin{figure*} 
\centering 
\includegraphics[width=14.0cm, angle=-0]{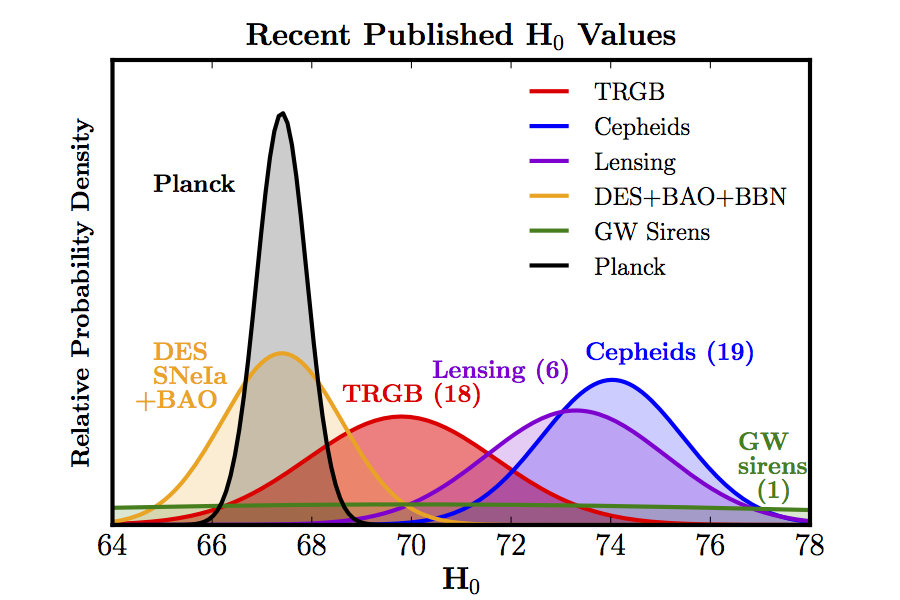}
\caption{\small The relative probability density distributions for several recent determinations of \ho (CMB: Planck Collaboration 2018; DES SNeIa + BAO: Macaulay et al. 2018 ;  TRGB: Freedman et al. 2019;  Lensing: Wong et al. 2019; Cepheids: Riess et al. 2019; Gravitational Wave Sirens: Abbott et al. 2017).}
\label{fig:Ho_summary}
\end{figure*}

We adopt the average of our measured values from \S\ref{sec:zero_points} of A$_I^{LMC}$ = 0.165 $\pm$ 0.02 mag. Based on the calibration for \hstacs F814W from Freedman et al. (2019), applied to a sample of nearby galaxies with TRGB distances, and tied to the Carnegie Supernova Project distant sample of \sne, we determine a slightly revised value of H$_0$ = 69.6 $\pm$ 0.8 ($\pm$1.1\% stat) $\pm$ 1.7 (±2.4\% sys) km s$^{-1}$ Mpc$^{-1}$, (a difference of 0.23\%). We note that despite the criticisms leveled by Y19, our value of H$_0$ agrees to within 1-$\sigma$ with their quoted value of  H$_0$ = 72.4  $\pm$ 1.9. In Figure \ref{fig:Ho_summary}, we show our value of \ho compared to other recently published determinations in the literature. Our purpose in this paper is not to undertake a detailed assessment of current \ho values. At this time, we note simply that the Hubble `tension' issue remains.

\section{Summary and Conclusions}
In this paper we have introduced and described in detail a new methodology for the simultaneous 
determination of true distance moduli and {\it total} line-of-sight reddenings for TRGB 
stars, self-consistently using multi-wavelength data for the TRGB stars themselves. 
We have used a set of stars in the outer main body of 
the LMC, combined with the recently published high-accuracy and high-precision 
distance to the LMC, to calibrate the TRGB method at five independent 
wavelengths ranging from the optical ($VI$) to the near infrared ($JHK$).

The multi-wavelength calibration resulting from the LMC analysis alone is given in \S\ref{sec:zero_points}. We find an absolute $I$-band magnitude for the TRGB of M$_I^{TRGB}$ = -4.047 $\pm$ 0.022 (stat) $\pm$ 0.039 (sys); and for F814W, M$_{F814W}^{TRGB}$ = -4.054 $\pm$ 0.022 (stat) $\pm$ 0.039 (sys). Independent consistency checks on this calibration have also been presented. The first is a calibration through TRGB stars in the SMC, using  a geometric distance to the SMC based on five DEBs. This calibration gives an $I$-band zero point of  $M_I =$~-4.09 $\pm$ 0.03 (stat) $\pm$ 0.05 (sys)~mag. The second is a near-infrared calibration of using a composite set of Galactic globular clusters with a range of (spectroscopically-determined) metallicities from [Fe/H] = -2.2 to -0.7 dex.  Without any additional free parameters, we set the zero-point calibration adopting the new DEB distance modulus to 47~Tuc, tied to the relative distances from DCA90. This gives an $I$-band zero point of  $M_I = $ -4.06 $\pm$ 0.05 (stat) $\pm$ 0.08 (sys)~mag. Each of these independent checks result in zero points consistent with our adopted value of our primary calibration using the LMC alone.

We find a value for the LMC TRGB extinction of A$_I$ = 0.160 $\pm$ 0.024 mag using the nearby galaxy, IC 1613, for the differential reddening comparison. This value agrees well with the independent value of A$_I$ = 0.169 $\pm$ 0.014 mag, using OGLE-III photometry for the SMC. This photometry suffers less from crowding/blending effects than that of Zaritsky et al. (2002), and leads to an improved estimate of the LMC TRGB extinction than that obtained by F19.  Our measurement technique has the advantage of self-consistently being determined by and for the TRGB stars themselves. Our (combined) value of A$_I$ = 0.165 $\pm$ 0.02 mag lies within one sigma of the mean for other recent estimates of the LMC extinction. It differs by 0.005 mag from that adopted in Freedman et al. (2019).

Adopting the \hstacs calibration from Freedman et al. (2019), applied to a sample of nearby galaxies with TRGB distances, and tied to the Carnegie Supernova Project distant sample of \sne, we determine a value of H$_0$ = 69.6 $\pm$ 0.8 ($\pm$1.1\% stat) $\pm$ 1.7 (±2.4\% sys) km s$^{-1}$ Mpc$^{-1}$.

We note that the measurement of the extinction  can continue to be improved with future study. We  have already obtained new high-resolution optical data using the 6.5-meter Magellan Baade telescope for several nearby galaxies, in order to further improve the accuracy of our LMC reddening estimate. 

Results published recently by Y19 differ from ours for several reasons. The first point of divergence is that we use the OGLE-III photometry of the SMC to derive the LMC extinction, rather than attempting, as Y19 do, to correct the problematic photometry of Zaritsky et al. (2002). Second, Y19 adjust the F19 result for single-band extinction alone, neglecting the concomitant, iterative corrections for reddening, which must be applied (as discussed in detail in \S\ref{sec:red_method}). Third, Y19 apply a theoretical correction for metallicity effects to the Freedman et al. (2019) result for the IC 1613 calibration, neglecting the fact that the method outlined in Freedman et al., and again in more detail in this paper, already explicitly accounts for metallicity. The methodology employed in Y19 differs significantly from the analysis process that we have explicated here and applied in Freedman et al.; and in major respects their interpretation of the method is simply in error. We conclude that our direct measurement of the LMC TRGB extinction {\it based on the TRGB stars themselves} is to be preferred over a(n arbitrary) value determined for a different population of stars, and which is systematically low compared to other recent studies in the published literature (see Figure \ref{fig:AI_values}). Despite these differences, we note that the Freedman et al. (2019) and Y19 H$_o$ values agree to within 1-$\sigma$.

An accurate measure of the extinction to TRGB stars in the LMC will remain an important (but not the only) component of the calibration and application of the TRGB method to the determination of H$_o$. Future improvements will come from higher precision multi-wavelength photometry for LMC TRGB stars, and from new multi-wavelength photometry and Gaia DR3 parallaxes to their counterparts in the halo of the Milky Way. Future Gaia parallaxes for Milky Way globular clusters will also strengthen the Galactic calibration of the TRGB. We note that the TRGB method, as applied to the determination of \ho, is relatively new compared to the application of the Cepheid distance scale or cosmic microwave background measurements. Additional calibrators for the TRGB method will be forthcoming from \hst and eventually with the {\it James Webb Space Telescope}. The high precision measured for the TRGB, and the ability to work in the isolated, low-reddening halos of galaxies of all morphological types, will be critical for breaking the impasse in current measurements of \ho.

\section{Acknowledgements} We thank Andre Udalski and collaborators for making 
their Magellanic Cloud surveys publicly available. We  thank Lars Bildsten 
for several enlightening discussions on the TRGB at a number of recent meetings. 
Early comments and suggestions from Adam Riess, regarding the optical calibration of 
the LMC stars, are gratefully acknowledged. We also thank the Padova Group, and especially Leo 
Giradi for making their stellar evolution code, PARSEC, publicly and easily available for use 
on the web. We thank the {\it Observatories of the Carnegie Institution for Science} and the 
{\it University of Chicago} for their support of our long-term research into 
the calibration and determination of the expansion rate of the Universe. Support for this work 
was provided in part by NASA through grant number HST-GO-13691.003-A from the {it Space Telescope 
Science Institute}, which is operated by AURA, Inc., under NASA contract NAS 5-26555. 
M.G.L. is supported by a grant from the National Research Foundation (NRF) of Korea, funded by the Korean Government (NRF-2019R1A2C2084019).
Partial support for this work was provided by NASA through Hubble Fellowship grant \#51386.01  to R.L.B. by the Space Telescope Science Institute, which is operated by the Association of  Universities for Research in Astronomy, Inc., for NASA, under contract NAS 5-26555.
This publication made use of data products from the Two Micron All Sky Survey, which is a joint project of the University of Massachusetts and the Infrared Processing and Analysis Center/California Institute of Technology, funded by the National Aeronautics and Space Administration and the National Science Foundation. We also made use of the NASA/IPAC Extragalactic Database (NED) which is operated by the Jet Propulsion Laboratory, California Institute of Technology, under contract with the National Aeronautics and Space Administration. And, finally, we thank the referee for many constructive suggestions and points of clarification that helped to improve this paper.

\vfill\eject
\section{References}
\medskip

\noindent
Abbott, B. P., Abbott, R. Abbott, T. D., Nature, 2017, 551, 85 

\noindent
Anand, G.,S., Tully, R.B., Rizzi, L., et al. 2019, ApJL, 872, L4

\noindent
Arenou, F., Luri, X., Babusiaux, C., et al. 2018, A\&A, 616, 17

\noindent
Bellazzini, M., Ferraro, F.R. \&  Pancino, E. 2001, ApJ, 556, 635 

\noindent
Bellazzini, M. 2008, MmSAI, 79, 440

\noindent
Bildsten, L., Paxton, B., Moore, K., et al. 2012, ApJL, 744, 6

\noindent
Bressan, A., Marigo, P., Girardi, L., et al. 2012, MNRAS, 427, 127

\noindent
Chen, S.,  Richer, H.,  Caiazzo, I.,  \&  Heyl, J. 2018, ApJ, 867, 132
 
\noindent
Cutri, R.M., Skrutskie, M.F., van Dyk, S., et al. 2006, 
{\it ``Explanatory Supplement to the 2MASS All Sky Data Release 
and Extended Mission Products''}
(Electronic version available on-line at: https://irsa.ipac.caltech.edu/data/2MASS/docs/releases/allsky/doc/explsup.html)

\noindent
da Costa, G.S. \& Armandroff, T.E. 1990, AJ, 100, 162 (DCA90)

\noindent
Dalcanton, J. J., Williams, B. F., Seth, A. C., et al., 2009, ApJS, 183, 67

\noindent
Dalcanton, J. J., Williams, B. F., Melbourne, J. L., et al., 2012, ApJS, 198, 6

\noindent
Fitzpatrick, E., 1999, PASP, 111, 63

\noindent
Freedman, W.L. 1988, ApJ, 326, 691

\noindent
Freedman, W.L. 1991, ApJ, 372, 455

\noindent
Freedman, W.L., Madore, B.F., Hatt, D., et al. 2019, ApJ, 882, 34

\noindent
Graczyk, D., Pietrzyński, G., Thompson, I.B., et al. 2014, ApJ, 780, 59

\noindent
Girardi, L. \& Salaris, M. 2001, MNRAS, 323, 109

\noindent
Gorski, M., Pietrzyński, G., Gieren, W., et al. 2016, AJ, 151, 167

\noindent
Gorski, M., Zgirski, B., Pietrzynski, G. et al. 2020, arXiv:2001.08242 

\noindent
Hatt, D., Beaton, R.L. Freedman, W.L.,  et al. 2017, ApJ, 845, 146

\noindent
Held, G.  Jozsa, G., Serra, P., et al. 2011, A\&A, 526, 118

\noindent
Hoffmann, S.L. \& Macri, L.M. 2015, AJ, 149, 183 

\noindent
Humphreys, E.M.L., Reid, M.J., Moran, J.M., et al. 2013, ApJ, 775, 13

\noindent
Ibata, R. A., Bellazzini, M, Malhan, K. et al.  2019, Nat. 
Astron., 3, 667

\noindent
Jacyszyn-Dobrzeniecka, A.~M., Skowron, D.~M., Mr{\'o}z, P., et al.\ 2016, \actaa, 66, 149

\noindent
Jang, I.-S. \& Lee, M.G. 2017a, ApJ, 835, 28

\noindent
Jang, I.-S. \& Lee, M.G. 2017b, ApJ, 836, 74

\noindent
Joshi, Y. C.  \& Panchal, A.  2019, A\&A 628, A51

\noindent
Lee, M. G., Freedman, W. L. \& Madore, B. F. 1993 ApJ, 417, 553

\noindent
Kaluzny, J., Thompson, I.B., Rozycka, M., et al. 2013, AJ, 145, 43 (M4)

\noindent
Kaluzny, J., Thompson, I.B., Dotter, A., et al. 2014, Acta Astron., 64, 11 (M55)

\noindent
Kaluzny, J., Thompson, I.B., Dotter, A., et al. 2013, 
AJ, 150, 155 (NGC~6362)

\noindent
Kiss, L.L \& Bedding, T.R. 2003, MNRAS, 343, L79

\noindent
Lee, M.-G., Freedman, W.L. \& Madore, B.F. 1993, ApJ, 417, 553 

\noindent
Makarov, D., Makarova, L., Rizzi, L., et al. 
2006, AJ, 132, 2729

\noindent
Madore, B.F., Freedman, W.L., Hatt, D., et al. 2018, ApJ, 858, 11

\noindent
Mager, V., Madore, B.F. \& Freedman, W.L., et al. 2008, 
ApJ, 689, 721 (Mag08)

\noindent
Marigo, P., Girardi, L., Bressan, A., et al., 2017, ApJ, 835, 77

\noindent
McQuinn, K.B.W., Boyer, M., Skillman, E.D., et al. 2019, ApJ, 880, 63

\noindent
Mosser, B., Dziembowski, W.A., Belkacem, K., et al. 2013, A\&A, A137

\noindent
Mould, J.R., Clementini, G. \& Da Costa, G. 2019, PASA, 36, 1 

\noindent
Mould, J. \& Sakai, S. 2008, ApJL, 686, 75

\noindent
Mould, J. \& Sakai, S. 2009, ApJ, 697, 996

\noindent
Macaulay, E., Nichol, R. C.   Bacon, D. et al.  
2018,  arXiv:1811.02376v2 

\noindent
Paczyński, B. 1997, Extragalactic Distance Scale STScI Symposium, eds. M. Livio, M. Donahue, and N. Panagia, Cambridge University Press, p. 273

\noindent
Pawlak, M., 2016, MNRAS, 547, 4323

\noindent
Planck Collaboration 2018, Aghanim, N., Akrami, Y., 
et al. 2018, arXiv:1807.06209
     
\noindent
Pietrzyński, G., Graczyk, D., Gallenne, A., et al. 2019, 
Nature, 567, 200

\noindent 
Riess, A. G., Macri, L. M., Hoffmann, S. L., et al. 2016, ApJ, 826, 56

\noindent 
Riess, A. G., Casertano, S., Yuan, W., et al. 2019, ApJ, 876, 85

\noindent
Rizzi, L., Tully, R.B., Makarov, D., et al. 2007, ApJ, 661, 815

\noindent
Salaris, M. \& Cassisi, S. 1997, MNRAS, 289, 406

\noindent
Sakai, S., Ferrarese, L., Kennicutt, R. C., Jr., \& Saha, A. 
2004, ApJ, 608, 42

\noindent
Schlafly, E.F. \& Finkbeiner, D.P. 2011, ApJ, 737, 103

\noindent
Schlegel, D.J., Finkbeiner, D.P \& Davis, M.C. 1998, ApJ, 500, 525

\noindent
Scowcroft, V., Freedman, W.L., Madore, B.F., et al. 2013, 
ApJ, 773, 106  

\noindent
Scowcroft, V., Freedman, W.L., Madore, B.F., et al. 2013, 
ApJ, 773, 106  

\noindent
Serenelli, A., Weiss, A., Cassisi, S., et al. 2017, A\&A, 606, 33

\noindent
Thompson, I.B., Udalski, A., Dotter, A., et al. 2020, MNRAS, eprint 2001.01481

\noindent
Udalski, A., Szymanski, M.K., Soszynski, I., et al. 2008, AcA, 58, 69	

\noindent
Ulaczyk, K., Symanski, M.K., Udalski, A., et al. 2012, AcA, 62, 247

\noindent
Valenti, E. Ferraro, F. R., Origlia, L. 2004a, MNRAS, 351, 1204

\noindent
Valenti, E. Ferraro, F. R., Origlia, L. 2004b, MNRAS, 354, 820

\noindent
Wielgorski, P. Pietrzyński, G., Gieren, W. et al. 2017, ApJ, 842, 116

\noindent
Williams, B.F., Dalcanton, J.J., Seth, A. C., et al. 2009, AJ, 137, 419

\noindent
Yuan, W., Riess, A. G., Macri, L.M., et al., 2019, ApJ, 886, 61 (Y19)

\noindent
Zaritsky, D., Harris, J., Thompson, I., et al. 2002, AJ, 123, 855

\noindent
Zaritsky, D., Harris, J., Thompson, I., et al. 2004, AJ, 128, 1606

\vfill\eject
\appendix
\section{Comments on the Yuan et al. (2019) Paper}
\label{app:appendix_Yuan}

After reverse-engineering the calibration plot presented in Freedman et al. (2019), Yuan et al. (2019; hereafter Y19) concluded that the results of our paper were incorrect. Here, we show that there are a number of misunderstandings and incorrect assumptions made in the Y19 paper. For example, in their critique of our results based on a differential analysis of the LMC reddening with respect to the SMC and IC~1613, they (back)-apply corrections to the F19 result, leading them to erroneous conclusions. We now list and discuss these issues in detail below.

\begin{enumerate}
    \item {\it The SMC: Use of the Zaritsky et al. (2002) data }
    
    Y19 noted that the OGLE-III photometry for the SMC is less subject to blending issues than that available to Freedman et al. (2019), which was based on the survey of Zaritsky et al. (2002; hereafter Z02). As seen earlier in \S \ref{sec:red_method}, we have now analyzed the OGLE-III photometry, and emphasize that the Y19 criticism of the SMC reddening (as being due to blending issues) is no longer applicable to the results in this current paper. Indeed, as shown in their Figure 5, there is no apparent trend of the photometric offsets with local number density.  We note that the difference in reddening, resulting from adopting the OGLE-III data over the Z02 data for the SMC, amounts to only +0.005~mag, and falls within the one-sigma uncertainty quoted by Freedman et al. (2019).                            
    
    \item {\it Y19 ``correction" to the Z02 SMC data}
    
    We note that the re-analysis by Y19 for the SMC data {\it is based on the Z02 photometry}, which they have 
    demonstrated are less accurate. Rather than use the better OGLE-III photometry, Y19  correct the Z02 results for blending effects, and then back-correct the Freedman et al. (2019) results. We point out that in doing this `back-correction', Y19 correct  for a magnitude effect (the extinction), but do not iteratively and simultaneously correct for the color (reddening) term that {\bf must} also be accounted for (see the discussion in \S\ref{sec:red_method}  and Figures  \ref{fig:schematic_ext+red}, and \ref{fig:ext_sim}; without taking these effects into account, the extinction and reddening terms will be underestimated.) We further note their blending correction may or may not be applicable to the actual stars used in the Freedman et al. analysis. We conclude that the Y19 `back-corrected' result for the SMC extinction value based on the Z02 data is incorrect.
    
    \item {\it IC 1613: Y19 statement about the ``V" data }
    
    Y19 conclude that they cannot make use of the data for IC 1613, used here and in Freedman et al. (2019), for a photometry comparison. They state that ``the IC 1613 detections are presented in Hatt et al. (2017)  where ground-based $V$ photometry is calibrated to a redder ACS filter, F606W." We wish to be clear here, that as outlined in \S\ref{sec:red_method}, we do not make use of such a transformation for our derived fit to the extinction law. The optical data that enter into the derivation of the extinction fit are zeroed to the $VI$ photometry from Hatt et al. (2017), based on the well-calibrated set of Stetson photometry (http://www.cadc-ccda.hia-iha.nrc-cnrc.gc.ca/en/community/STETSON/standards/). In Freedman et al. and Hatt et al. we compare these ground-based data to the F814W data for IC~1613,  and find agreement to $\pm$0.02 mag (1\%). To briefly summarize, we fit an extinction law to a multi-wavelength ($VIJHK$) set of data to determine A$_I$ for the LMC, which we then apply to the $I$-band TRGB magnitude. The transformation to the \hstacs F814W system is done only in the last step, when we calibrate the TRGB magnitude. The data referred to in Hatt et al. are not used and are not relevant for the analysis for determining the reddening described here.

    \item {\it IC 1613: Y19 double correction for metallicity }
    
    Y19 cite McQuinn et al. (2019; hereafter M19) stating that ``the use of a single slope to `rectify' the TRGB for its color dependence at different metallicities produces differences in the TRGB of $\sim$0.04 mag for the difference in metallicity between -2 to -1 dex. Quoting directly, Y19 state that they 
    ``naively use the comparison in McQuinn et al. (2019) for the same SFH and $\Delta[Fe/H]$ (from -2 to -1~dex)"  to illustrate the impact of rectifying the TRGB colors on the extinction estimate. These corrections amount to 0.026, 0.033 and 0.038 mag, making the TRGB stars fainter in $J$, $H$ and $K$, respectively; and amounting to -0.019 and -0.014~mag offsets in $V$ and $I$, respectively, making stars at those wavelengths brighter. After making these corrections, Y19 find a value for A$_I^{LMC}$ = 0.10 mag ``in agreement with earlier determinations." We have discussed the (broad) range of earlier published reddening determinations in \S\ref{sec:reddening}. However, we point out that the correction applied by Y19 for metallicity is simply incorrect, as illustrated by both the theoretical isochrones shown in Figures \ref{fig:VIK} and \ref{fig:VIJHK}, and by the empirical color-magnitude diagrams shown in Figures \ref{fig:LMC_CMDs} and \ref{fig:SMC_1613_CMDs}. The well-known metallicity-tracking trends with color for RGB stars are {\it directly} exhibited in the CMDs. Y19 have thus apparently applied corrections for metallicity effects twice.
    
    We summarize the results of Figure \ref{fig:VIJHK} once again here: The TRGB stars at $I$, for example, map uniquely into their respective positions in the CMDs at other colors, and they do so in accordance with their metallicities as reflected and calibrated by their colors. As {\it expected}, the most metal-rich stars are fainter at $V$, and brighter at $K$.  These other wavebands do not have a flat distribution with magnitude; there is a well-defined {\it run} or sequence of magnitude with color {\it that is not arbitrary}. Using a self-consistent sample of {\it precisely the same RGB stars} defined in the $I$-band CMD, one can make use of the (very fortunate) fact that the optical ($V$ and $I$), and the near-infared  ($JHK$) bandpasses exhibit distinctly different behaviors with metallicity. The opposite signs of the optical/near-IR behavior resulting from  metallicity differences is in contrast to that of extinction, which is a steadily decreasing function of inverse wavelength. The different dependencies of metallicity and extinction with wavelength thus allows the two effects to be separated, and the extinction can be independently measured, as demonstrated in Figure \ref{fig:schematic_ext+red}, and detailed in \S\ref{sec:schematic_ext+red}.


\item{\it LMC Reddening}
  
Y19 argue that the extinction estimates from Haschke et al. (2011) are 
the best available estimate of the LMC reddening. The recent results of Gorski et al. (2020) dispute that claim, arguing for reddenings that are 0.06~mag higher in $E(B-V)$ than the Haschke et al. calibration, which is based on red clump (RC) stars and RR Lyrae stars. Unlike the case for TRGB stars, the evolutionary tracks for RC stars are well-known to be strong functions of both age and metallicity (see, for example, Girardi \& Salaris (2001) and Williams et al. (2009)).
Finally, as  discussed by Joshi \& Panchal (2019), and in agreement with Gorski et al., it is found that the Haschke extinction estimates are systematically lower than all other estimates reviewed in their summary. For completeness, we note also that a higher value for the RC star extinction is also found in the earlier analysis by Pawlak (2016).
At the time of their study, Jang \& Lee (2017) had no  
direct estimate of the TRGB extinction and  
adopted the Haschke et al. extinction value, with its originally published uncertainty of 0.07 mag.  

\end{enumerate}

\end{document}